\documentclass[acmsmall,screen]{acmart}
\usepackage[shortlabels]{enumitem}
\usepackage{fancybox, graphicx}
\usepackage{color}
\usepackage{listings}
\usepackage{tcolorbox}
\usepackage{balance}
\usepackage{tabularx}
\usepackage{multirow}
\usepackage{xspace}
\usepackage{tikz}
\usepackage{textcomp}
\usepackage{booktabs}
\usetikzlibrary{patterns}
\usepackage{url}

\definecolor{darkblue}{rgb}{0.0, 0.0, 0.55}  

\usepackage[normalem]{ulem}
\usepackage[utf8]{inputenc}
\usepackage{amsmath}  
\usepackage{graphicx}
\usepackage{comment}
\usepackage{multirow}
\usepackage{enumitem}
\usepackage{array}
\usepackage{pifont}
\usepackage{xcolor}
\usepackage{balance}
\usepackage{mathtools}
\usepackage{xspace}
\usepackage{url}
\usepackage{eucal}
\usepackage{amsfonts}
\usepackage{color}
\usepackage{booktabs}
\usepackage{bbding}
\usepackage{float}

\usepackage{graphicx}
\usepackage{hyperref}
\usepackage{url}
\usepackage{multirow}
\usepackage{comment}
\usepackage{colortbl}
\usepackage{graphicx}
\usepackage{subcaption}
\usepackage[linewidth=0.1pt]{mdframed}
\usepackage[linesnumbered,ruled,lined]{algorithm2e}

\usepackage{wrapfig,lipsum,booktabs}
\usepackage{subfloat}
\usepackage[export]{adjustbox}

\usepackage{graphicx}
\usepackage{adjustbox}
\usepackage{forest}
\usepackage{tikz}
\usepackage{pifont}

\usepackage{hyperref}
\usepackage{hyperxmp}

\AtBeginDocument{%
  \providecommand\BibTeX{{%
    \normalfont B\kern-0.5em{\scshape i\kern-0.25em b}\kern-0.8em\TeX}}}

\newcommand{\mrevised}[1]{\textcolor{black}{#1}}

\pdfpagewidth=\paperwidth
\pdfpageheight=\paperheight

\makeatletter
\newcommand{\showfontsize}{\f@size{} pt}
\newcommand\usemm[1]{%
  \strip@pt\dimexpr0.3514598\dimexpr #1\relax\relax mm%
}
\newcommand\usein[1]{%
  \strip@pt\dimexpr0.013837\dimexpr #1\relax\relax in%
}
\makeatother

\usepackage{graphicx}
\usepackage{hyperref}
\usepackage{url}
\usepackage{multirow}
\usepackage{comment}
\usepackage{colortbl}
\usepackage{graphicx}
\usepackage{tcolorbox}
\usepackage{arydshln}
\usepackage{algpseudocode}  
\usepackage{amsmath}  
\usepackage{caption} 
\usepackage{algpseudocode}
\usepackage{float}
\usepackage{soul}

\usepackage{xcolor}
\usepackage{enumitem}

\usepackage{amsmath}

\definecolor{dark-red}{RGB}{255,0,0}
\definecolor{dark-green}{RGB}{0,200,0}

\lstdefinelanguage{java-pretty}
{
  language=java,
  numbers=left,
  frame=shadowbox,
  rulesepcolor= \color{red!20!green!20!blue!20},
  basicstyle=\footnotesize\ttfamily,
  numberstyle=\scriptsize,
  breaklines=true,
  columns=fullflexible,
  xleftmargin=16pt,
  showstringspaces=false,
  keywordstyle=\color{blue}\bfseries,
  stringstyle=\color{javared},
  commentstyle=\color{javagreen},
  morecomment=[s][\color{javadocblue}]{/**}{*/},
}
\colorlet{punct}{red!60!black}
\definecolor{background}{HTML}{EEEEEE}
\definecolor{delim}{RGB}{20,105,176}
\colorlet{numb}{magenta!60!black}
\lstdefinelanguage{json}{
    basicstyle=\normalfont\ttfamily,
    numbers=left,
    numberstyle=\scriptsize,
    stepnumber=1,
    numbersep=8pt,
    showstringspaces=false,
    breaklines=true,
    frame=lines,
    literate=
     *{:}{{{\color{punct}{:}}}}{1}
      {,}{{{\color{punct}{,}}}}{1}
      {\{}{{{\color{delim}{\{}}}}{1}
      {\}}{{{\color{delim}{\}}}}}{1}
      {[}{{{\color{delim}{[}}}}{1}
      {]}{{{\color{delim}{]}}}}{1},
}

\lstdefinelanguage{json-pretty}
{
  language=json,
  numbers=left,
  frame=shadowbox,
  rulesepcolor= \color{red!20!green!20!blue!20},
  basicstyle=\footnotesize\ttfamily,
  numberstyle=\scriptsize,
  breaklines=true,
  columns=fullflexible,
  xleftmargin=16pt,
  showstringspaces=false,
  keywordstyle=\color{blue}\bfseries,
  stringstyle=\color{javared},
  commentstyle=\color{javagreen},
  morecomment=[s][\color{javadocblue}]{/**}{*/},
}

\lstdefinelanguage{diff}{
    basicstyle=\ttfamily\small,
    morecomment=[f][\color{diffstart}]{@@},
    morecomment=[f][\color{javagreen}]{+\ },
    morecomment=[f][\color{javared}]{-\ },
  }

\newcommand{\InputWithSpace}[1]{\bgroup\def\arraystretch{1.15}\input{#1}\egroup}

\setcopyright{acmcopyright}
\copyrightyear{2024}
\acmYear{2024}
\acmDOI{XXXXXXX.XXXXXXX}

\begin{document}

\title{Large Language Model for Vulnerability Detection and Repair: Literature Review and the Road Ahead}

\author{Xin Zhou}
\affiliation{%
  \institution{Singapore Management University}
  \country{Singapore}
}
\email{xinzhou.2020@phdcs.smu.edu.sg}
\orcid{0000-0002-4558-0622}

\author{Sicong Cao}
\affiliation{%
    \institution{Yangzhou University}
  \country{China}
}
\email{Dx120210088@yzu.edu.cn}
\orcid{0000-0003-3688-4437}

\author{Xiaobing Sun}
\affiliation{%
    \institution{Yangzhou University}
  \country{China}
}
\email{xbsun@yzu.edu.cn}
\orcid{0000-0001-5165-5080}

\author{David Lo}
\affiliation{%
  \institution{Singapore Management University}
  \country{Singapore}
}
\email{davidlo@smu.edu.sg}
\orcid{0000-0002-4367-7201}

\renewcommand{\shortauthors}{Zhou et al.}



\begin{abstract}
The significant advancements in Large Language Models (LLMs) have resulted in their widespread adoption across various tasks within Software Engineering (SE), including vulnerability detection and repair. Numerous studies have investigated the application of LLMs to enhance vulnerability detection and repair tasks. Despite the increasing research interest, there is currently no existing survey that focuses on the utilization of LLMs for vulnerability detection and repair. In this paper, we aim to bridge this gap by offering a systematic literature review of approaches aimed at improving vulnerability detection and repair through the utilization of LLMs. The review encompasses research work from leading SE, AI, and Security conferences and journals, encompassing 43 papers published across 25 distinct venues, along with 15 high-quality preprint papers, bringing the total to 58 papers. By answering three key research questions, we aim to (1) summarize the LLMs employed in the relevant literature, (2) categorize various LLM adaptation techniques in vulnerability detection, and (3) classify various LLM adaptation techniques in vulnerability repair. Based on our findings, we have identified a series of limitations of existing studies. Additionally, we have outlined a roadmap highlighting potential opportunities that we believe are pertinent and crucial for future research endeavors.
\end{abstract}

\maketitle

\newcommand{\XSpace}[1]{}

\makeatletter
\newenvironment{btHighlight}[1][]
{\begingroup\tikzset{bt@Highlight@par/.style={#1}}\begin{lrbox}{\@tempboxa}}
{\end{lrbox}\bt@HL@box[bt@Highlight@par]{\@tempboxa}\endgroup}

\newcommand\btHL[1][]{%
  \begin{btHighlight}[#1]\bgroup\aftergroup\bt@HL@endenv%
}
\def\bt@HL@endenv{%
  \end{btHighlight}%
  \egroup
}
\newcommand{\bt@HL@box}[2][]{%
  \tikz[#1]{%
    \pgfpathrectangle{\pgfpoint{1pt}{0pt}}{\pgfpoint{\wd #2}{\ht #2}}%
    \pgfusepath{use as bounding box}%
    \node[anchor=base west, fill=yellow!30,outer sep=0pt,inner xsep=1pt, inner ysep=0pt, rounded corners=0pt, minimum height=\ht\strutbox+1pt,#1]{\raisebox{1pt}{\strut}\strut\usebox{#2}};
  }%
}
\makeatother
\lstdefinestyle{Highlight}{
    moredelim=**[is][\btHL]{`}{`},
    moredelim=**[is][{\btHL[fill=orange!50]}]{´}{´},
    moredelim=**[is][{\btHL[fill=red!50]}]{@}{@},
}

\newcommand{\wholedata}{{\textit{complete data}}}
\newcommand{\singledata}{{\textit{single-line data}}}

\section{Introduction}

A software vulnerability refers to a flaw or weakness in a software system that can be exploited by attackers. Recently, the number of software vulnerabilities has increased significantly~\cite{security_report2}, affecting numerous software systems. 
To mitigate these issues, researchers have proposed methods for automatic detection
and repair of identified vulnerabilities.
However, traditional techniques, such as rule-based detectors or program analysis-based repair tools, encounter challenges due to high false positive rates~\cite{shahriar2012mitigating} and their inability to work for diverse types of vulnerabilities~\cite{zhou2024out}, respectively.

Recently, Large Language Models (LLMs) pre-trained on large corpus have demonstrated remarkable effectiveness across various natural language and software engineering tasks~\cite{Hou2023LargeLM}.
Given their recent success, researchers have proposed various LLM-based approaches to improve automated vulnerability detection and repair, demonstrating promising outcomes for both detection and repair tasks~\cite{steenhoek2023empirical,zhou2024out}.
LLM-based approaches for vulnerability detection and repair are increasingly attracting attention due to their potential to automatically learn features from known vulnerabilities and find/fix unseen ones. 
Furthermore, LLMs have the potential to utilize rich knowledge acquired from large-scale pre-training to enhance vulnerability detection and repair.

Despite the increasing research interest in utilizing LLMs for vulnerability detection and repair, to the best of our knowledge, no comprehensive literature review has yet summarized the state-of-the-art approaches, identified the limitations of current studies, and proposed future research directions in this field. To effectively chart the most promising path forward for research on the utilization of LLM techniques in vulnerability detection and repair, we conducted a \textbf{Systematic Literature Review (SLR)} to bridge this gap, providing valuable insights to the community. In this paper, we collected \textbf{58} primary studies over the last 6 years (2018-2024). We then summarized the LLMs used in these studies, classified various techniques for adapting LLMs, and discussed the limitations of existing research. Finally, we proposed a roadmap outlining future research directions and opportunities in this area.

\begin{figure*}[t]
  \centering 
  \begin{adjustbox}{width=1.0
  \columnwidth}
    \begin{forest}
for tree={
    rounded corners,
    child anchor=west,
    parent anchor=east,
    grow'=east,  
    text width=4cm,%
    draw=darkblue,
    anchor=west,
    node options={align=center},
    edge path={
      \noexpand\path[\forestoption{edge}]
        (.child anchor) -| +(-5pt,0) -- +(-5pt,0) |-
        (!u.parent anchor)\forestoption{edge label};
    },
    where n children=0{text width=7cm}{}
  },
  [LLM-based methods for vulnerability detection and repair
    [Background and Preliminary \S\ref{sec:background}],
    [Review Methodology \S\ref{sec:method}],
    [LLMs Used for Vulnerability Detection and Repair \S\ref{RQ1}],
    [How to Adapt LLMs for Vulnerability Detection \S\ref{RQ2}
        [Fine-tuning \S\ref{sec_detection:fine-tuning}
            [Data-centric Innovations ~\cite{DBLP:conf/icse/YangWLW23,DBLP:conf/kbse/WenWGWLG23,DBLP:conf/scam/KuangYZTY23,DBLP:journals/corr/abs-2403-18624}],
            [Combination of LLM with Program Analysis ~\cite{liu2024pre,DBLP:conf/scam/PengCZTLH23,wang2023combining,DBLP:journals/tse/ZhangLHXL23,tran2024detectvul,DBLP:conf/internetware/WengQLLC24}],
            [Combination of LLM with Other Deep Learning Modules ~\cite{DBLP:journals/jss/TangTBZF23,DBLP:conf/infocom/ZiemsW21,DBLP:conf/internetware/JiangSGWW0024,yang2024security}],
            [Domain-specific Pre-training ~\cite{DBLP:conf/ijcnn/HanifM22,liu2024pre,ni2023distinguishing,wang2023combining,DBLP:journals/corr/abs-2403-18624,DBLP:journals/corr/abs-2311-04109} ],
            [Causal Learning ~\cite{mahbubur2023towards}],
            [Default Fine-tuning ~\cite{DBLP:conf/msr/FuT22,DBLP:conf/acsac/ThapaJACPN22,DBLP:journals/ese/FuTLKNPG24,DBLP:conf/icse/SejfiaDSM24,DBLP:conf/uss/RisseB24,DBLP:conf/icse/SteenhoekRJL23,DBLP:conf/raid/0001DACW23,DBLP:conf/icse/0027WZLZX0024,DBLP:conf/icse/CroftBK23,DBLP:conf/msr/LeDB24,zhou2024comparison,DBLP:journals/corr/abs-2401-17010}],
        ],
        [Prompt Engineering \S\ref{sec_detection:PromptEngineering}
            [Zero-shot Prompting ~\cite{fu2023chatgpt, DBLP:journals/corr/abs-2401-15468, DBLP:conf/issre/PurbaGRC23,DBLP:journals/corr/abs-2311-16169,DBLP:journals/corr/abs-2308-12697,zhou2024comparison,ni2024learning,DBLP:journals/corr/abs-2404-03994} ],
            [Few-shot Prompting ~\cite{ni2024learning,zhou2024comparison} ],
        ],
        [Retrieval Augmentation \S\ref{sec_detection:RetrievalAugmentation} ~\cite{DBLP:conf/dsc/LiuLGG23,DBLP:journals/corr/abs-2401-15468,DBLP:journals/corr/abs-2406-11147,DBLP:journals/corr/abs-2404-15596}
        ],
    ],
    [How to Adapt LLMs for Vulnerability Repair \S\ref{RQ3}
        [Fine-tuning \S\ref{sec_repair:fine-tuning}
            [Data-Centric Innovations ~\cite{zhou2022spvf, zhou2024out, DBLP:journals/ejwcn/WeiBWLYSL23,DBLP:journals/corr/abs-2402-13291}],
            [Model-Centric Innovations ~\cite{fu2023vision,DBLP:journals/corr/abs-2401-03374}],
            [Domain-specific Pre-training ~\cite{DBLP:journals/tse/ChenKM23,DBLP:journals/tse/ChiQLZY23,zhang2023pre,zhou2024out}],
            [Reinforcement Learning ~\cite{islam2024code}],
            [Default Fine-tuning ~\cite{DBLP:conf/sigsoft/FuTLNP22,DBLP:journals/ese/FuTLKNPG24}],
        ],
        [Prompt Engineering \S\ref{sec_repair:PromptEngineering}
            [Zero-shot Prompting ~\cite{pearce2023examining,DBLP:conf/issta/WuJPLD0BS23,DBLP:journals/corr/abs-2302-01215,DBLP:journals/corr/abs-2402-17230,DBLP:journals/corr/abs-2405-04994}],
             [Few-shot Prompting ~\cite{fu2023chatgpt,nong2024automated}],
        ],
    ],
    [Characteristics of Datasets and Deployment  \S\ref{RQ4}
       [Data Characteristics \S\ref{DataCharacteristics}],
       [Deployment Strategies \S\ref{DeploymentStrategies}],
    ],
    [The Road Ahead \S\ref{ChaAndOpp}
       [Limitations \S\ref{Challenges}],
       [Roadmap \S\ref{Roadmap}],
    ],
  ]
\end{forest}
\end{adjustbox}
\caption{Structure of This Survey} 
\label{fig:structure}
\end{figure*}

\vspace{0.1cm}
\noindent\textbf{Related Literature Reviews.}
Researchers have undertaken a series of research endeavors concerning Machine Learning (ML) for source code vulnerability detection or repair~\cite{Hou2023LargeLM,ghaffarian2017software,lin2020software,wu2022code,zhang2023surveyrepair}.
Hou et al.~\cite{Hou2023LargeLM} conducted a systematic literature review on LLMs for SE. Similarly, Zhang et al.~\cite{zhang2023unifying} systematically reviewed the recent advancements in LLMs for SE. However, due to the extensive range of SE tasks to summarize, their review did not categorize the detailed utilization of LLMs in vulnerability detection and repair. In contrast, our literature review is more focused on vulnerability detection and repair.
Ghaffarian et al.~\cite{ghaffarian2017software} studied vulnerability analysis and discovery approaches published till 2016.
Lin et al.~\cite{lin2020software} reviewed vulnerability detection approaches utilizing deep learning till 2020.
Wu et al.~\cite{wu2022code} investigated vulnerability detection approaches published before May 2022. 
However, their focus was not on LLMs and primarily covered general ML models such as recurrent neural networks. In contrast, our literature review includes papers published until March 2024 and focuses on LLMs, encompassing 37 distinct LLMs utilized in 58 relevant studies.
Recently, Zhang et al.~\cite{zhang2023survey} reviewed learning-based automated program repair, which covered vulnerability repair as part of its study scope. In Different from Zhang et al., our literature review is more focused on LLMs and vulnerability repair: we included more recent studies, offered a detailed categorization of LLM usages in vulnerability repair, and discussed specialized future directions for vulnerability repair with LLMs.

In general, this study makes the following \textbf{contributions}:
\vspace{-\topsep}
\begin{itemize}[leftmargin=1em]
\item We present a systematic review of recent \textbf{58} primary studies focusing on the utilization of LLMs for vulnerability detection and repair.
\item 
We offer a comprehensive summary of the LLMs utilized in the relevant literature and categorize various techniques used to adapt LLMs to do the two tasks (vulnerability detection and repair).
\item We discuss the limitations of existing studies on using LLMs for vulnerability detection and repair and propose a roadmap outlining future research directions and opportunities.
\end{itemize}

\textbf{Survey Structure.} Figure~\ref{fig:structure} outlines the structure of this paper. Section~\ref{sec:background} provides the background information, while Section~\ref{sec:method} details the methodology. Section~\ref{RQ1} discusses the LLMs employed in vulnerability detection and repair. Sections~\ref{RQ2} and~\ref{RQ3} cover the relevant work on adapting and enhancing LLMs for vulnerability detection and repair, respectively. 
Section~\ref{RQ4} discusses the characteristics of datasets and deployment in LLM-based vulnerability detection and repair studies.
Finally, Section~\ref{ChaAndOpp} discusses the limitations of current studies and presents a roadmap for future research and opportunities.

\section{Background and Preliminaries}
~\label{sec:background}

\begin{figure*}[t]
  \centering
  \includegraphics[width=1\linewidth]{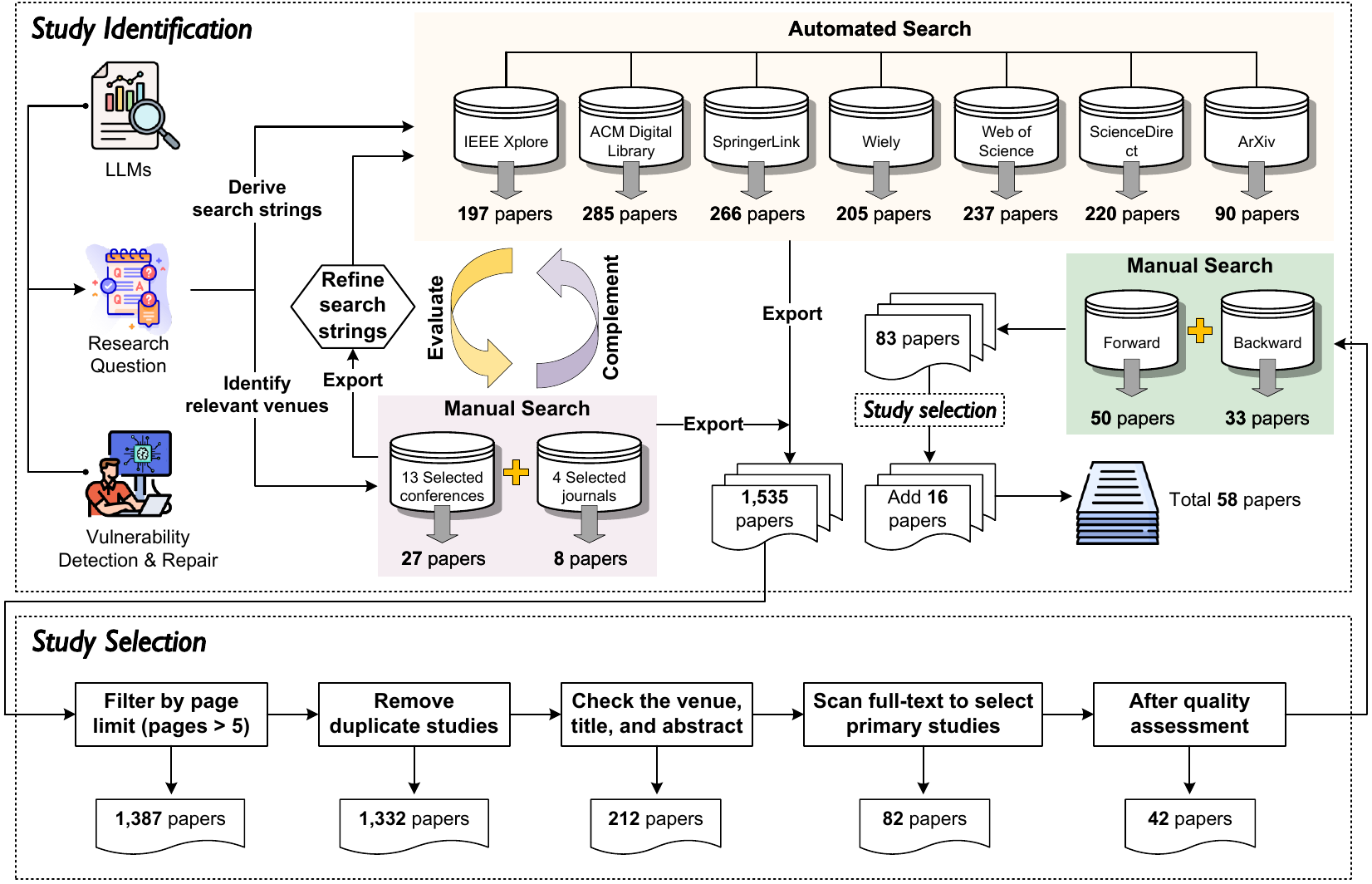}
\caption{Study Identification and Selection Process}
\label{search_framework}
\end{figure*}

\subsection{Vulnerability Detection/Repair Formulation}
In this study, we focus on source code vulnerability detection and repair. Here, we present the task formulations of these two tasks.  

\vspace{0.1cm}
\noindent\textbf{Vulnerability Detection.}
Vulnerability detection is typically framed as a binary classification task:
$X_i \to Y_i$. Specifically, given an input source code function $X_i$, a model predicts whether the input function is vulnerable ($Y_i=1$) or non-vulnerable ($Y_i=0$).

\vspace{0.1cm}
\noindent\textbf{Vulnerability Repair.}
Vulnerability repair studies using LLMs frame the task as a sequence-to-sequence problem: $X_i \to Y_i$. Specifically, given a vulnerable code snippet $X_i$, an LLM model generates the corresponding repaired code $Y_i$.

\vspace{-0.3cm}
\subsection{Large Language Models (LLMs)} 
The term Large Language Model (LLM) was introduced to distinguish language models based on their parameter size, specifically referring to large-sized pre-trained language models~\cite{zhao2023survey}. However, the literature lacks a formal consensus on the minimum parameter scale for LLMs~\cite{wang2023software}. In this paper, we adopt the LLM scope division and taxonomy introduced by Pan et al.~\cite{pan2023unifying} and categorize the mainstream LLMs into three groups according to their architectures: 1) encoder-only, 2) encoder-decoder, and 3) decoder-only LLMs.
We will provide a brief introduction to some representative LLMs for each category due to the space limit.

\vspace{0.1cm}
\noindent\textbf{Encoder-only LLMs. }
Encoder-only LLMs are a type of neural network architecture that utilizes only the encoder component of the Transformer model~\cite{devlin2018bert}. 
In the SE domain, examples of encoder-only LLMs include  CodeBERT~\cite{feng2020codebert}, GraphCodeBERT~\cite{guo2020graphcodebert}, CuBERT~\cite{Kanade2019LearningAE},  
VulBERTa~\cite{DBLP:conf/ijcnn/HanifM22}, CCBERT~\cite{zhou2023ccbert}, SOBERT~\cite{he2023representation}, and BERTOverflow~\cite{tabassum2020code}.

\vspace{0.1cm}
\noindent\textbf{Encoder-decoder LLMs.}
Encoder-decoder LLMs integrate both the encoder and decoder modules of the Transformer model~\cite{vaswani2017attention}. The encoder processes the input sentence, while the decoder generates the target output text/code.
Prominent examples of encoder-decoder LLMs include PLBART~\cite{ahmad2021unified}, T5~\cite{raffel2020exploring}, CodeT5~\cite{wang2021codet5}, UniXcoder~\cite{DBLP:conf/acl/GuoLDW0022}, and NatGen~\cite{DBLP:conf/sigsoft/ChakrabortyADDR22}.

\vspace{0.1cm}
\noindent\textbf{Decoder-only LLMs.}
Decoder-only LLMs exclusively utilize the decoder module of the Transformer model to generate the target output text/code. 
The GPT series, including GPT-2~\cite{radford2019language}, GPT-3~\cite{brown2020language}, GPT-3.5~\cite{openai2022gpt}, and GPT-4~\cite{openai2023gpt4}, stand as prominent implementations of this model series.
Additionally, in the SE domain, there are numerous decoder-only LLMs specialized for code as well. Examples include CodeGPT~\cite{lu2021codexglue}, Codex~\cite{chen2021evaluating}, Polycoder~\cite{DBLP:conf/pldi/Xu0NH22}, Incoder~\cite{DBLP:conf/iclr/FriedAL0WSZYZL23}, 
CodeGen series~\cite{DBLP:conf/iclr/NijkampPHTWZSX23,DBLP:journals/corr/abs-2305-02309}, Copilot~\cite{github2021copilot}, Code Llama~\cite{Meta2023Codellama}, and StarCoder~\cite{Li2023StarCoderMT}.

\section{Review Methodology}
~\label{sec:method}

\subsection{Research Question}

In this paper, we focus on investigating four Research Questions (RQs). The first three RQs mainly focus on identifying which LLMs are used and how they are adapted for vulnerability detection and repair:
\begin{itemize}[leftmargin=1em] 
\item \textbf{RQ1: What LLMs have been utilized to solve vulnerability detection and repair tasks?} 
\item \textbf{RQ2: How are LLMs adapted for vulnerability detection?} 
\item \textbf{RQ3: How are LLMs adapted for vulnerability repair?} 
\end{itemize} 
The first three RQs pertain primarily to the model construction phase. Generally, the pipeline of a learning-based method can be divided into three phases: 1) data preparation, 2) model construction, and 3) deployment. We also investigate the characteristics of the datasets and deployment designs in LLM-based vulnerability detection and repair studies: 
\begin{itemize}[leftmargin=1em] 
\item \textbf{ RQ4: What are the characteristics of the datasets and deployment strategies in LLM-based vulnerability detection and repair studies?} 
\end{itemize}

\vspace{-0.3cm}
\subsection{Search Strategy}
As shown in Fig.~\ref{search_framework}, following the guide by Zhang et al.~\cite{DBLP:journals/infsof/ZhangBT11}, our initial step is to identify primary studies to answer the research questions (RQs) above.
Because the first LLM (i.e., BERT~\cite{devlin2018bert}) was introduced in 2018, our search focused on papers published from 2018 onwards (i.e., from January 2018, to March 2024).
Next, we identified the top peer-reviewed and influential conference and journal venues in the domains of SE, AI, and Security. We included 13 conferences (ICSE, ESEC/FSE, ASE, ISSTA, CCS, S\&P, USENIX Security, NDSS, AAAI, IJCAI, ICML, NIPS, ICLR) and 4 journals (TOSEM, TSE, TDSC, TIFS). 
After the manual searching, we identified 35 papers that were relevant to our research objectives.

In addition to manually searching primary studies from top-tier venues, we also conducted an automated search across 7 popular databases, including IEEE Xplore~\cite{IEEEXplore}, ACM Digital Library~\cite{ACMDigital}, SpringerLink~\cite{SpringerLink}, Wiely~\cite{Wiely}, ScienceDirect~\cite{ScienceDirect},
Web of Science~\cite{WebofScience} and arXiv~\cite{arXiv}.
The search string used in the automated search is crafted from the relevant papers identified in the manual search. Please kindly check the complete set of search keywords in our online appendix~\cite{appendix} due to the limited space.
After conducting the automatic search, we collected 1,500 relevant studies with the automatic search from these 7 popular databases.

\vspace{-0.3cm}
\subsection{Study Selection} 

\vspace{0.1cm}
\noindent\textbf{Inclusion and Exclusion Criteria.}
After the paper collection, we conducted a relevance assessment according to the following inclusion and exclusion criteria:

\vspace{-\topsep}
\vspace{0.05cm}
\begin{itemize}[leftmargin=1.5em]
\item[{\textcolor[RGB]{0,128,0}{\Checkmark}}] \emph{The paper must be written in English.}
\item[{\textcolor[RGB]{0,128,0}{\Checkmark}}] \emph{The paper must have an accessible full text.}
\item[{\textcolor[RGB]{0,128,0}{\Checkmark}}] \emph{
The paper must be a peer-reviewed full research paper published either in a conference proceeding or a journal. \footnote{For preprint papers released on arXiv in 2023 or 2024 that are not yet published due to their recent release, we will retain them when checking the venues and will perform a manual quality check to decide whether to include them in our study.}}
\item[{\textcolor[RGB]{0,128,0}{\Checkmark}}] \emph{The paper must adopt LLM techniques to solve source code vulnerability detection or repair.}
\item[{\textcolor[RGB]{209,26,66}{\XSolidBrush}}] \emph{The paper has less than 5 pages.}
\item[{\textcolor[RGB]{209,26,66}{\XSolidBrush}}] \emph{Books, keynote records,  panel summaries, technical reports, theses, tool demos papers, editorials, or venues not subject to a full peer-review process.}
\item[{\textcolor[RGB]{209,26,66}{\XSolidBrush}}] \emph{The paper is a literature review or survey.}
\item[{\textcolor[RGB]{209,26,66}{\XSolidBrush}}] \emph{Duplicate papers or similar studies authored by the same authors.}
\item[{\textcolor[RGB]{209,26,66}{\XSolidBrush}}] \emph{The paper does not utilize LLMs, e.g., using graph neural networks.}
\item[{\textcolor[RGB]{209,26,66}{\XSolidBrush}}] \emph{The paper mentions LLMs only in future work or discussions rather than using LLMs in the approach.}
\item[{\textcolor[RGB]{209,26,66}{\XSolidBrush}}] \emph{The paper does not involve source code vulnerability detection or repair tasks.}
\end{itemize}

\vspace{-\topsep}
\vspace{0.02cm}
In the first phase, by filtering out short papers (exclusion criteria 1) and deduplication (exclusion criteria 4), the total number of included papers was reduced to 1,332. In the second phase, we manually examined the venue, title, and abstracts of the papers, and the total number of included papers declined to 212. 
Please kindly note that, in this step, we retain preprint papers released on arXiv in 2023 or 2024 that are not yet published due to their recent release.
Books, keynote records,  panel summaries, technical reports, theses, tool demo papers, editorials, literature reviews, or survey papers were also discarded in this phase (exclusion criteria 2-3). 
In the third phase, we manually read the full text of the paper to remove irrelevant papers. 
Specifically, the vulnerability detection or repair papers, which do not utilize LLMs but other methods, e.g., graph neural networks (GNN) or recurrent neural networks (RNN), were dropped (exclusion criteria 5).
We also excluded studies that do not focus on source code vulnerability, such as those on binary code, protocols, or network communication.
Furthermore, we removed studies that just discussed LLM as an idea or future work (exclusion criteria 6). We also removed the papers focusing on other tasks rather than vulnerability detection or repair, such as vulnerable data generation, vulnerability assessment, etc. (exclusion criteria 7).
After the third phase, we identify 82 primary studies directly relevant to our research topic.

\begin{figure}[t] 
\huge
\centering 
\includegraphics[width=0.7\linewidth]{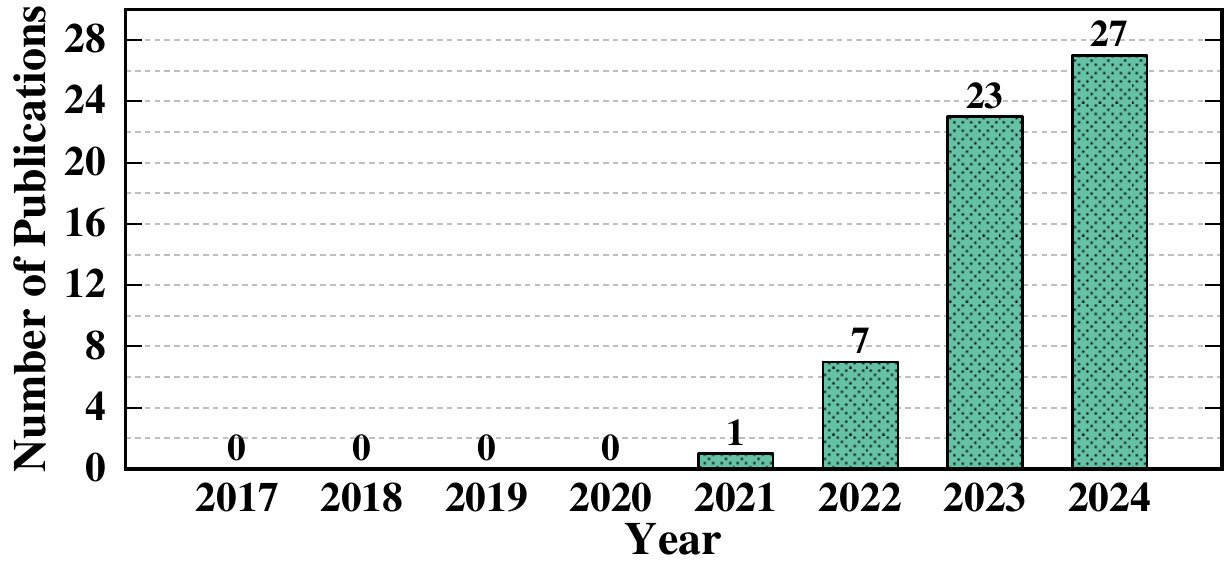} 
\vspace{-0.3cm}
\caption{Distribution of Publications per Year}
\vspace{-0.7cm}
\label{PubPerYear} 
\end{figure}

\vspace{0.05cm}
\noindent\textbf{Quality Assessment.}
To prevent biases introduced by low-quality studies, we formulated five Quality Assessment Criteria (QAC):
\begin{itemize}[leftmargin=*]
    \item \emph{QAC$_1$:} Was the study published in a prestigious venue?
    \item \emph{QAC$_2$:} Does the study make a contribution to the academic or industrial community?
    \item \emph{QAC$_3$:} Does the study provide a clear description of the workflow and implementation of the proposed approach? 
    \item \emph{QAC$_4$:} Are the experiment details, including datasets, baselines, and evaluation metrics, clearly outlined?
    \item  \emph{QAC$_5$:} Do the findings from the experiments strongly support the main arguments presented in the study?
\end{itemize}
We employed a scoring system ranging from 0 to 3 (poor, fair, good, excellent) for each quality assessment criterion. Following the manual assignment of scores, we selected papers with total scores reaching 12 (80\% of the maximum possible score).
Please note that for preprint papers released on arXiv in 2023 or 2024 that are not yet published, their score for the ``venue'' criterion is 0. However, if their total score reaches 12, we still consider them in our study.
After this quality assessment, we obtained 42 papers.
Among them, 37 papers were published, and 5 high-quality preprint papers were included after passing the quality assessment.

\vspace{0.05cm}
\noindent\textbf{Forward and Backward Snowballing.}
To avoid omitting any possibly relevant work during our manual and automated search process, we also performed lightweight backward and forward snowballing ~\cite{DBLP:conf/ease/Wohlin14}. This involved reviewing both the references cited in our selected 42 primary studies and the publications that cited these studies. As a supplement, we gathered 83 more papers and repeated the entire study selection process, including filtering, deduplication, and quality assessment, which resulted in the identification of 16 additional papers. \emph{Thus, we obtained a final set of \textbf{58} papers to study.}

\vspace{-0.2cm}
\subsection{Data Extraction and Analysis}\label{DataEx}
Fig.~\ref{PubPerYear} illustrates the distribution of selected primary studies across each year. The earliest relevant study we identified was published in 2021~\cite{DBLP:conf/infocom/ZiemsW21}. Subsequently, interest in exploring LLMs for vulnerability detection and repair has steadily increased, peaking in 2024, with 46.6\% of the total studied papers. 
The growing number of papers on these topics indicates increasing research interest in leveraging LLMs for vulnerability detection and repair.
We also analyzed the publication venues of the included studies. ICSE emerges as the predominant conference venue for LLM studies on vulnerability detection and repair, contributing 20.7\% of the total number of studies. Other notable venues are TSE (5.2\%), FSE (3.4\%), EMSE (3.4\%), ISSTA (3.4\%), MSR (3.4\%), and TOSEM (1.7\%).

\section{RQ1: What LLMs have been utilized?}\label{RQ1}

\begin{figure}[b] 
\centering 
\includegraphics[width=0.7\linewidth]{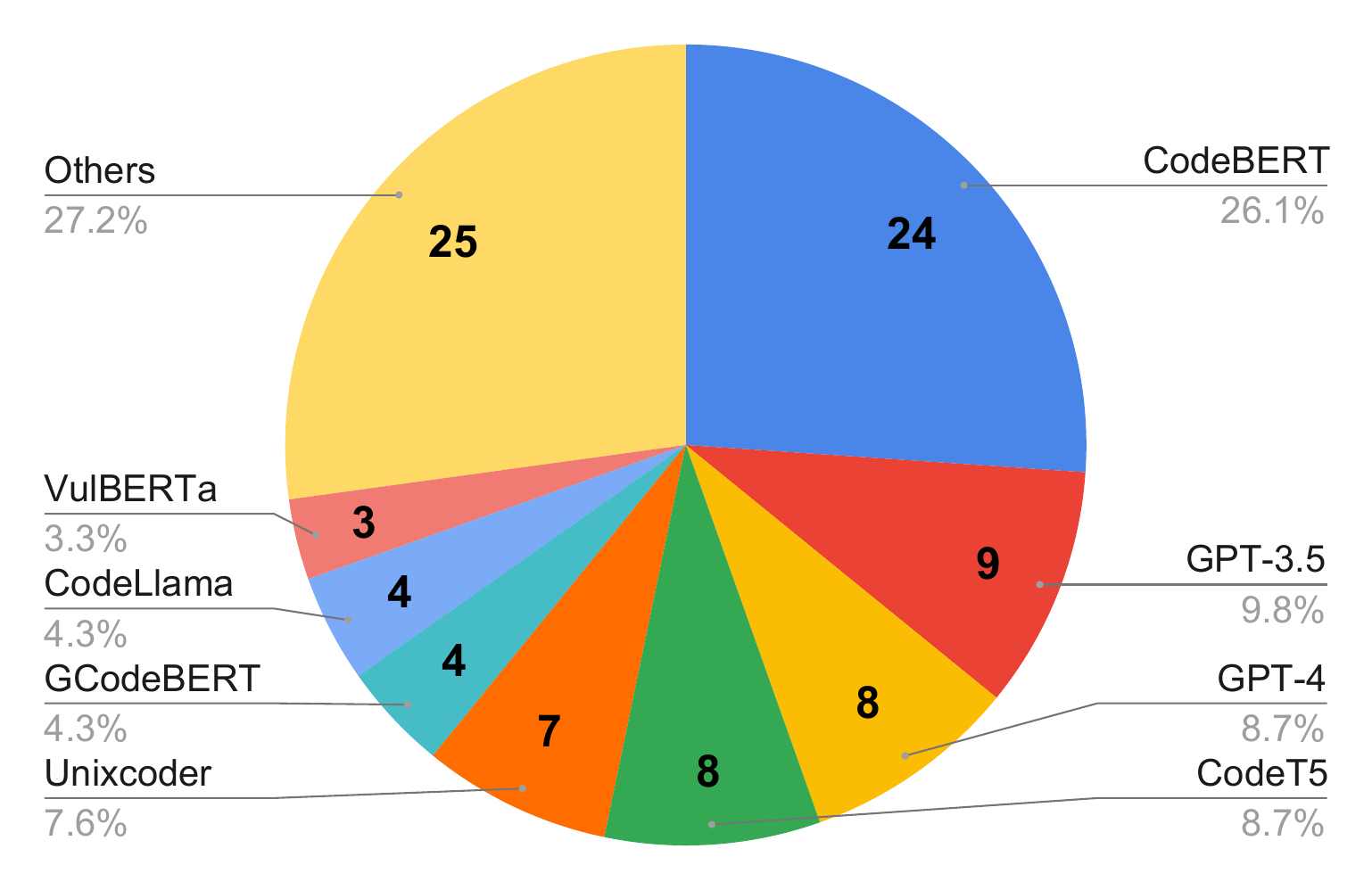} 
\caption{Distribution of LLMs for Vulnerability Detection}
\label{RQ1A} 
\end{figure}

\begin{figure}[t] 
\centering 
\includegraphics[width=0.7\linewidth]{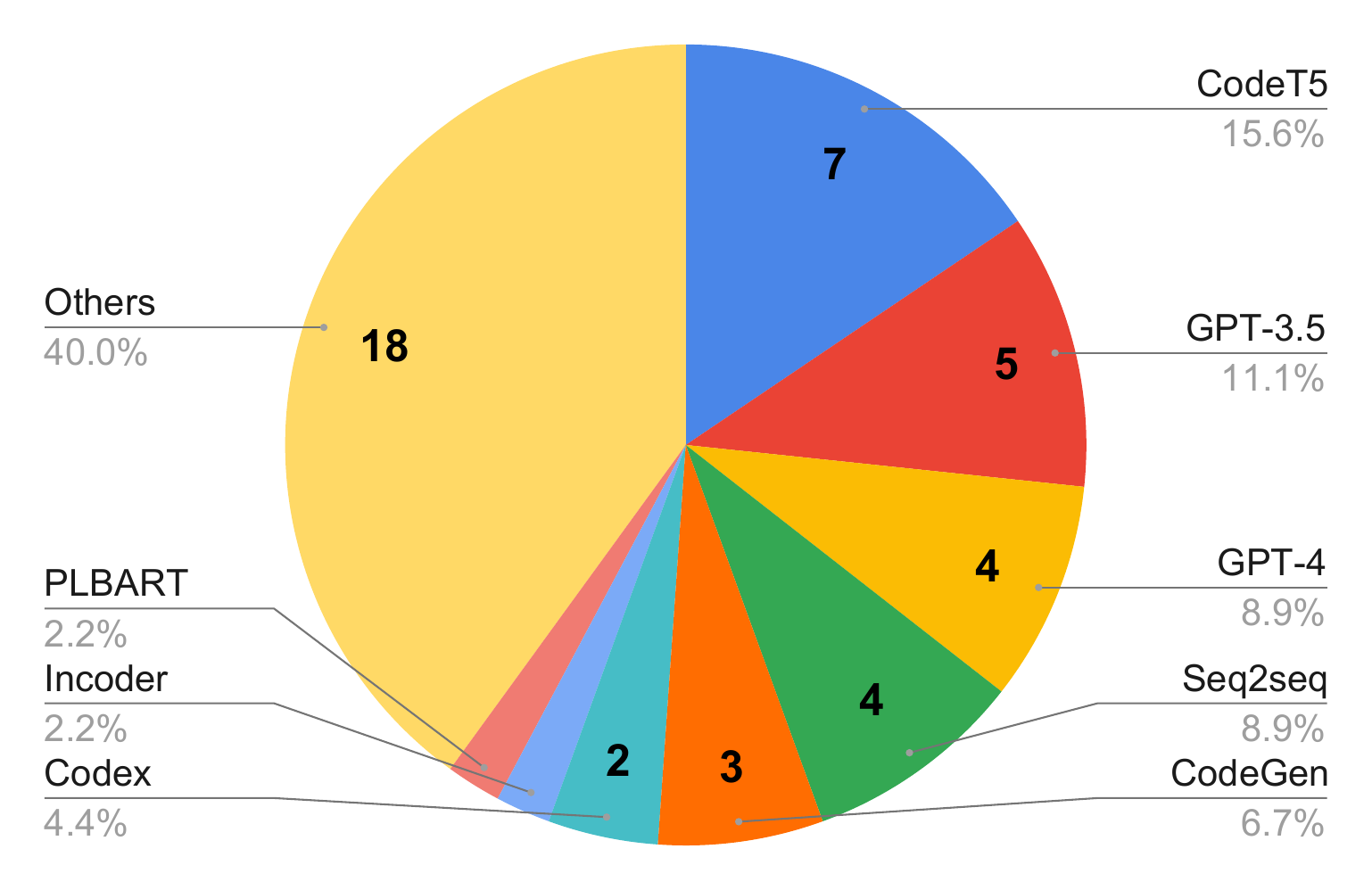} 
\vspace{-0.2cm}
\caption{Distribution of LLMs for Vulnerability Repair}
\label{RQ1B} 
\end{figure}

In general, out of the 58 included studies, we identified 37 distinct LLMs that have been utilized. 

\vspace{0.2cm}
\noindent\textbf{LLMs Used for Vulnerability Detection.}
Fig. \ref{RQ1A} illustrates the distribution of the LLMs utilized for vulnerability detection. Please note that we count the occurrence of LLMs used in each study, and a single study may utilize multiple LLMs.
CodeBERT~\cite{feng2020codebert} emerges as the predominant LLM in addressing vulnerability detection to date, representing 26.1\% (24/92) of all the use of LLMs in the included studies. GPT-3.5 becomes the second most frequently studied model, accounting for 9.8\% (9/92).
Regarding the categories of LLMs, encoder-only LLMs comprise 47.8\% (44/92) of the use of LLMs in the included studies. Following that, decoder-only LLMs account for 23.9\%  (22/92), and encoder-decoder LLMs constitute 9.8\% (9/92). Lastly, commercial LLMs with undisclosed architectures, such as GPT-3.5 and GPT-4, account for 18.5\% (17/92) of the LLM usages.

\vspace{0.2cm}
\noindent\textbf{LLMs Used for Vulnerability Repair.}
Fig.~\ref{RQ1B} illustrates the distribution of LLMs utilized for vulnerability repair.
Unlike the detection task, CodeT5~\cite{wang2021codet5} emerges as the predominant LLM in addressing vulnerability repair to date, representing 15.6\% (7/45) of the use of LLMs in the included studies. GPT-3.5 becomes the second most frequently studied model, accounting for 11.1\% (5/45). Following closely, GPT-4 and domain-specific pre-trained Seq2Seq Transformers become the third most frequently studied model, accounting for 8.9\% (4/45) of the use of LLMs. 
Regarding the categories of LLMs, decoder LLMs comprise 31.1\% (14/45) of LLMs used. Following that, encoder-decoder LLMs and encoder-only LLMs constitute 26.7\% (12/45)  and 6.7\% (3/45) of the use of LLMs, respectively.  Lastly, commercial LLMs with undisclosed architectures, such as GPT-3.5 and GPT-4, account for 35.6\% (16/45) of the LLM usages.

\begin{table*}[]
\caption{Top 10 LLMs Used in Vulnerability Detection and Repair}
\vspace{-0.1cm}
\centering
\resizebox{0.85\width}{!}{%
\begin{tabular}{l|lll|lll}
\hline
\multirow{2}{*}{\textbf{\begin{tabular}[c]{@{}l@{}}Usages\\ Ranking\end{tabular}}} & \multicolumn{3}{c|}{\textbf{Vulnerability Detection}} & \multicolumn{3}{c}{\textbf{Vulnerability Repair}} \\ \cline{2-7} 
 & \multicolumn{1}{l|}{\textbf{LLM}} & \multicolumn{1}{l|}{\textbf{Structure}} & \textbf{Size} & \multicolumn{1}{l|}{\textbf{LLM}} & \multicolumn{1}{l|}{\textbf{Structure}} & \textbf{Size} \\ \hline
1 & \multicolumn{1}{l|}{CodeBERT} & \multicolumn{1}{l|}{Encoder-only} & 125M & \multicolumn{1}{l|}{CodeT5} & \multicolumn{1}{l|}{Encoder-Decoder} & 220M \\ \hline
2 & \multicolumn{1}{l|}{GPT-3.5} & \multicolumn{1}{l|}{Unknown} & Unknown & \multicolumn{1}{l|}{GPT-3.5} & \multicolumn{1}{l|}{Unknown} & Unknown \\ \hline
3 & \multicolumn{1}{l|}{GPT-4} & \multicolumn{1}{l|}{Unknown} & Unknown & \multicolumn{1}{l|}{GPT-4} & \multicolumn{1}{l|}{Unknown} & Unknown \\ \hline
4 & \multicolumn{1}{l|}{CodeT5} & \multicolumn{1}{l|}{Encoder-Decoder} & 220M & \multicolumn{1}{l|}{Seq2seq} & \multicolumn{1}{l|}{Encoder-Decoder} & 60M \\ \hline
5 & \multicolumn{1}{l|}{UnixCoder} & \multicolumn{1}{l|}{Encoder-only} & 126M & \multicolumn{1}{l|}{CodeGen} & \multicolumn{1}{l|}{Decoder-only} & \begin{tabular}[c]{@{}l@{}}350M, 2.7B, \\ 6.1B, 16.1B\end{tabular} \\ \hline
6 & \multicolumn{1}{l|}{GraphCodeBERT} & \multicolumn{1}{l|}{Encoder-only} & 125M & \multicolumn{1}{l|}{Codex} & \multicolumn{1}{l|}{Unknown} & 12.0B \\ \hline
7 & \multicolumn{1}{l|}{CodeLlama} & \multicolumn{1}{l|}{Decoder-only} & \begin{tabular}[c]{@{}l@{}}7B, 13B, \\ 34B, 70B\end{tabular} & \multicolumn{1}{l|}{Incoder} & \multicolumn{1}{l|}{Decoder-only} & 1.3B, 6.7B \\ \hline
8 & \multicolumn{1}{l|}{VulBERTa} & \multicolumn{1}{l|}{Encoder-only} & 125M & \multicolumn{1}{l|}{PLBART} & \multicolumn{1}{l|}{Encoder-Decoder} & 406M \\ \hline
9 & \multicolumn{1}{l|}{RoBERTa} & \multicolumn{1}{l|}{Encoder-only} & 125M & \multicolumn{1}{l|}{UnixCoder} & \multicolumn{1}{l|}{Encoder-only} & 126M \\ \hline
10 & \multicolumn{1}{l|}{BERT} & \multicolumn{1}{l|}{Encoder-only} & 109M & \multicolumn{1}{l|}{CodeGPT} & \multicolumn{1}{l|}{Decoder-only} & 124M \\ \hline
\end{tabular}
}
\label{table:top10-llm}
\end{table*}

In addition, Table~\ref{table:top10-llm} presents the architecture and model sizes of the top 10 most commonly used LLMs in vulnerability detection and repair. The architectures and model sizes of GPT-3.5 and GPT-4, along with the architecture of Codex, remain undisclosed due to their commercial status. As shown in Table~\ref{table:top10-llm}, the LLMs widely utilized for vulnerability detection are predominantly lightweight models, with many having sizes under or equal to 126 million parameters. Most of these models are encoder-only architectures. In contrast, the models commonly used for vulnerability repair tend to be larger, with several exceeding 1 billion parameters, such as CodeGen~\cite{DBLP:conf/iclr/NijkampPHTWZSX23} and Incoder~\cite{DBLP:conf/iclr/FriedAL0WSZYZL23}. The majority of these models have more than 126 million parameters and are primarily decoder-only and encoder-decoder architectures.

\begin{tcolorbox}[left=4pt,right=4pt,top=2pt,bottom=2pt,boxrule=0.5pt]
\textbf{Answer to RQ1:} 
Our analysis indicates that, to date, encoder-only LLMs have dominated vulnerability detection, while commercial LLMs and decoder-only LLMs have been prominent in vulnerability repair. 
\end{tcolorbox}

\section{RQ2: How Are LLMs Adapted for Vulnerability Detection?}\label{RQ2}

\begin{figure}[b]
  \centering
  \includegraphics[width=0.5\linewidth]{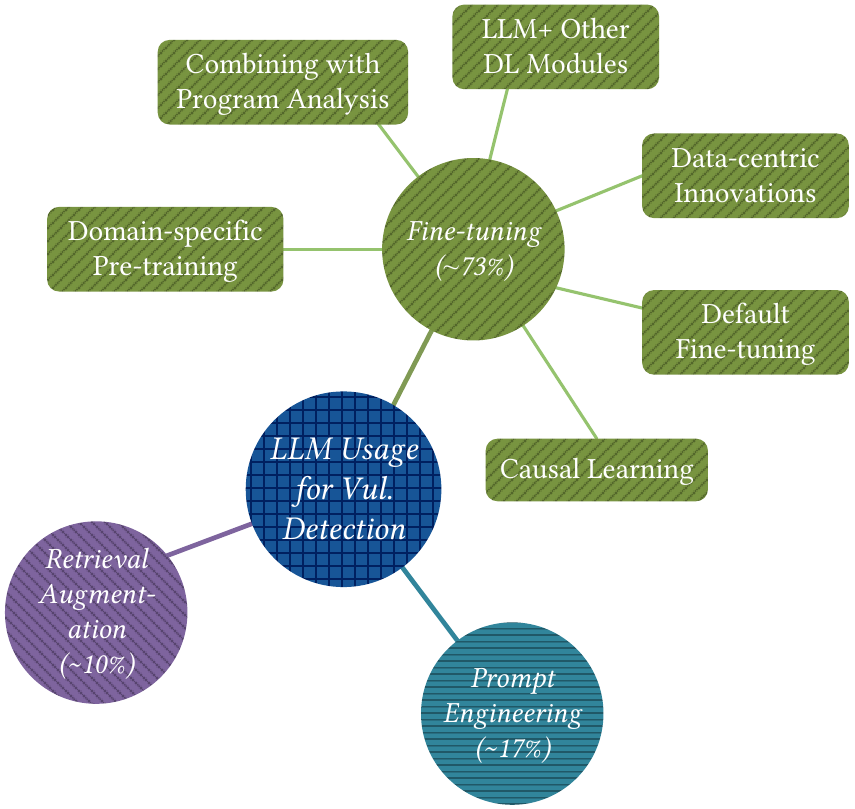}
\caption{Adaptation Techniques of LLMs for Vul. Detection}
\label{ExpTec}
\end{figure}

In RQ2, we shift our focus to examining the specific adaptation techniques of LLMs in vulnerability detection. 
Regarding the usage of LLMs, we summarize \textbf{\emph{three major categories}} from the included studies: \textbf{1) fine-tuning}~\cite{devlin2018bert}, which updates the parameters of LLMs using a labeled dataset, \textbf{2) prompt engineering}~\cite{brown2020language}, which designs prompts to guide LLMs in generating relevant responses without updating parameters, 
and \textbf{3) retrieval augmentation (RAG)}~\cite{DBLP:conf/nips/LewisPPPKGKLYR020}, which integrates knowledge from retrieval systems into the LLMs' context to improve their performance, also without changing LLMs' parameters.
As depicted in Fig.~\ref{ExpTec}, 73\% of the studies utilize fine-tuning, while 17\% and 10\% employ prompt engineering and retrieval augmentation, respectively.

\begin{table*}[]
\caption{Existing LLM-based Methods for Vulnerability Detection}
\resizebox{0.5\width}{!}{%
\begin{tabular}{l|c|lllllcll|cc|cc}
\toprule
\multirow{2}{*}{} & \textbf{LLMs} & \multicolumn{8}{c|}{\textbf{LLM Adaptation Techniques}} & \multicolumn{2}{c|}{\textbf{Data}} & \multicolumn{2}{c}{\textbf{Deployment}} \\ \cline{2-14} 
 & \multicolumn{1}{c|}{\textbf{Size}} & \multicolumn{1}{c|}{\begin{tabular}[c]{@{}c@{}}\textbf{Data-}\\ \textbf{centric}\end{tabular}} & \multicolumn{1}{c|}{\begin{tabular}[c]{@{}c@{}}\textbf{LLM+} \\ \textbf{PA}\end{tabular}} & \multicolumn{1}{c|}{\begin{tabular}[c]{@{}c@{}}\textbf{LLM+} \\ \textbf{DL}\end{tabular}} & \multicolumn{1}{c|}{\begin{tabular}[c]{@{}c@{}}\textbf{Domain} \\ \textbf{Training}\end{tabular}} & \multicolumn{1}{c|}{\begin{tabular}[c]{@{}c@{}}\textbf{Causal} \\ \textbf{Learning}\end{tabular}} & \multicolumn{1}{c|}{\begin{tabular}[c]{@{}c@{}}\textbf{Default} \\ \textbf{Tuning}\end{tabular}} & \multicolumn{1}{c|}{\begin{tabular}[c]{@{}c@{}}\textbf{Prompt} \\ \textbf{Eng.}\end{tabular}} & \multicolumn{1}{c|}{\begin{tabular}[c]{@{}c@{}}\textbf{RAG}\end{tabular}} & \multicolumn{1}{c|}{\begin{tabular}[c]{@{}c@{}}\textbf{Input} \\ \textbf{Granularity}\end{tabular}} & \begin{tabular}[c]{@{}c@{}}\textbf{Data} \\ \textbf{Labeling}\end{tabular} & \multicolumn{1}{c|}{\begin{tabular}[c]{@{}c@{}}\textbf{Interact}\\\textbf{with Users}\end{tabular}} & \begin{tabular}[c]{@{}c@{}}\textbf{Integrated}\\ \textbf{to Workflow}\end{tabular} \\ \hline
Ziems et al.~\cite{DBLP:conf/infocom/ZiemsW21} & <1B & \multicolumn{1}{c|}{} & \multicolumn{1}{c|}{} & \multicolumn{1}{c|}{\textcolor{green}{\huge \ding{51}}} & \multicolumn{1}{c|}{} & \multicolumn{1}{c|}{} & \multicolumn{1}{c|}{} & \multicolumn{1}{c|}{} & \multicolumn{1}{c|}{} & \multicolumn{1}{c|}{Function} & synthetic & \multicolumn{1}{c|}{\textcolor{red}{\huge \ding{55}}} & \textcolor{red}{\huge \ding{55}} \\ \hline
Fu et al.~\cite{DBLP:conf/msr/FuT22} & <1B & \multicolumn{1}{c|}{} & \multicolumn{1}{c|}{} & \multicolumn{1}{c|}{} & \multicolumn{1}{c|}{} & \multicolumn{1}{c|}{} & \multicolumn{1}{c|}{\textcolor{green}{\huge \ding{51}}} & \multicolumn{1}{c|}{} & \multicolumn{1}{c|}{} & \multicolumn{1}{c|}{Line} & heuristic & \multicolumn{1}{c|}{\textcolor{red}{\huge \ding{55}}} & \textcolor{red}{\huge \ding{55}} \\ \hline
Hanif et al.~\cite{DBLP:conf/ijcnn/HanifM22} & <1B & \multicolumn{1}{c|}{} & \multicolumn{1}{c|}{} & \multicolumn{1}{c|}{} & \multicolumn{1}{c|}{\textcolor{green}{\huge \ding{51}}} & \multicolumn{1}{c|}{} & \multicolumn{1}{c|}{} & \multicolumn{1}{c|}{} & \multicolumn{1}{c|}{} & \multicolumn{1}{c|}{Function} & \begin{tabular}[c]{@{}c@{}}synthetic, \\ heuristic, \\partially \\ manually labeled\end{tabular} & \multicolumn{1}{c|}{\textcolor{red}{\huge \ding{55}}} & \textcolor{red}{\huge \ding{55}} \\ \hline
Thapa et al.~\cite{DBLP:conf/acsac/ThapaJACPN22} & <10B & \multicolumn{1}{c|}{} & \multicolumn{1}{c|}{} & \multicolumn{1}{c|}{} & \multicolumn{1}{c|}{} & \multicolumn{1}{c|}{} & \multicolumn{1}{c|}{\textcolor{green}{\huge \ding{51}}} & \multicolumn{1}{c|}{} & \multicolumn{1}{c|}{} & \multicolumn{1}{c|}{Function} & heuristic & \multicolumn{1}{c|}{\textcolor{red}{\huge \ding{55}}} & \textcolor{red}{\huge \ding{55}} \\ \hline
Fu et al.~\cite{fu2023chatgpt} & >10B & \multicolumn{1}{c|}{} & \multicolumn{1}{c|}{} & \multicolumn{1}{c|}{} & \multicolumn{1}{c|}{} & \multicolumn{1}{c|}{} & \multicolumn{1}{c|}{} & \multicolumn{1}{c|}{\textcolor{green}{\huge \ding{51}}} & \multicolumn{1}{c|}{} & \multicolumn{1}{c|}{Function} & heuristic & \multicolumn{1}{c|}{\textcolor{red}{\huge \ding{55}}} & \textcolor{red}{\huge \ding{55}} \\ \hline
Wen et al.~\cite{DBLP:conf/kbse/WenWGWLG23} & <1B & \multicolumn{1}{c|}{\textcolor{green}{\huge \ding{51}}} & \multicolumn{1}{c|}{} & \multicolumn{1}{c|}{} & \multicolumn{1}{c|}{} & \multicolumn{1}{c|}{} & \multicolumn{1}{c|}{} & \multicolumn{1}{c|}{} & \multicolumn{1}{c|}{} & \multicolumn{1}{c|}{Function} & \begin{tabular}[c]{@{}c@{}}heuristic,\\ partially \\ manually labeled\end{tabular} & \multicolumn{1}{c|}{\textcolor{red}{\huge \ding{55}}} & \textcolor{red}{\huge \ding{55}} \\ \hline
Purba et al.~\cite{DBLP:conf/issre/PurbaGRC23} & >1B & \multicolumn{1}{c|}{} & \multicolumn{1}{c|}{} & \multicolumn{1}{c|}{} & \multicolumn{1}{c|}{} & \multicolumn{1}{c|}{} & \multicolumn{1}{c|}{} & \multicolumn{1}{c|}{\textcolor{green}{\huge \ding{51}}} & \multicolumn{1}{c|}{} & \multicolumn{1}{c|}{Function} & heuristic & \multicolumn{1}{c|}{\textcolor{red}{\huge \ding{55}}} & \textcolor{red}{\huge \ding{55}} \\ \hline
Liu et al.~\cite{DBLP:conf/dsc/LiuLGG23} & >10B & \multicolumn{1}{c|}{} & \multicolumn{1}{c|}{} & \multicolumn{1}{c|}{} & \multicolumn{1}{c|}{} & \multicolumn{1}{c|}{} & \multicolumn{1}{c|}{} & \multicolumn{1}{c|}{} & \multicolumn{1}{c|}{\textcolor{green}{\huge \ding{51}}} & \multicolumn{1}{c|}{Function} & \begin{tabular}[c]{@{}c@{}}partially \\ manually labeled\end{tabular} & \multicolumn{1}{c|}{\textcolor{red}{\huge \ding{55}}} & \textcolor{red}{\huge \ding{55}} \\ \hline
Peng et al.~\cite{DBLP:conf/scam/PengCZTLH23} & <1B & \multicolumn{1}{c|}{} & \multicolumn{1}{c|}{\textcolor{green}{\huge \ding{51}}} & \multicolumn{1}{c|}{} & \multicolumn{1}{c|}{} & \multicolumn{1}{c|}{} & \multicolumn{1}{c|}{} & \multicolumn{1}{c|}{} & \multicolumn{1}{c|}{} & \multicolumn{1}{c|}{Line} & heuristic & \multicolumn{1}{c|}{\textcolor{red}{\huge \ding{55}}} & \textcolor{red}{\huge \ding{55}} \\ \hline
Kuang et al.~\cite{DBLP:conf/scam/KuangYZTY23} & <1B & \multicolumn{1}{c|}{\textcolor{green}{\huge \ding{51}}} & \multicolumn{1}{l|}{} & \multicolumn{1}{l|}{} & \multicolumn{1}{l|}{} & \multicolumn{1}{l|}{} & \multicolumn{1}{l|}{} & \multicolumn{1}{l|}{} &  & \multicolumn{1}{c|}{Function} & \begin{tabular}[c]{@{}c@{}}heuristic,\\ partially \\ manually labeled\end{tabular} & \multicolumn{1}{c|}{\textcolor{red}{\huge \ding{55}}} & \textcolor{red}{\huge \ding{55}} \\ \hline
Zhang et al.~\cite{DBLP:journals/tse/ZhangLHXL23} & <1B & \multicolumn{1}{l|}{} & \multicolumn{1}{c|}{\textcolor{green}{\huge \ding{51}}} & \multicolumn{1}{l|}{} & \multicolumn{1}{l|}{} & \multicolumn{1}{l|}{} & \multicolumn{1}{l|}{} & \multicolumn{1}{l|}{} &  & \multicolumn{1}{c|}{Function} & \begin{tabular}[c]{@{}c@{}}heuristic,\\ partially \\ manually labeled\end{tabular} & \multicolumn{1}{c|}{\textcolor{red}{\huge \ding{55}}} & \textcolor{red}{\huge \ding{55}} \\ \hline
Tang et al.~\cite{DBLP:journals/jss/TangTBZF23} & <1B & \multicolumn{1}{l|}{} & \multicolumn{1}{l|}{} & \multicolumn{1}{c|}{\textcolor{green}{\huge \ding{51}}} & \multicolumn{1}{l|}{} & \multicolumn{1}{l|}{} & \multicolumn{1}{l|}{} & \multicolumn{1}{l|}{} &  & \multicolumn{1}{c|}{Function} & \begin{tabular}[c]{@{}c@{}}partially \\ manually labeled\end{tabular} & \multicolumn{1}{c|}{\textcolor{red}{\huge \ding{55}}} & \textcolor{red}{\huge \ding{55}} \\ \hline
Ni et al.~\cite{ni2023distinguishing} & <1B & \multicolumn{1}{l|}{} & \multicolumn{1}{l|}{} & \multicolumn{1}{l|}{} & \multicolumn{1}{c|}{\textcolor{green}{\huge \ding{51}}} & \multicolumn{1}{l|}{} & \multicolumn{1}{l|}{} & \multicolumn{1}{l|}{} &  & \multicolumn{1}{c|}{Function} & \begin{tabular}[c]{@{}c@{}}heuristic,\\crowd-sourced\end{tabular} & \multicolumn{1}{c|}{\textcolor{red}{\huge \ding{55}}} & \textcolor{red}{\huge \ding{55}} \\ \hline
Zhou et al.~\cite{DBLP:journals/corr/abs-2401-15468} & >10B & \multicolumn{1}{l|}{} & \multicolumn{1}{l|}{} & \multicolumn{1}{l|}{} & \multicolumn{1}{l|}{} & \multicolumn{1}{l|}{} & \multicolumn{1}{l|}{} & \multicolumn{1}{c|}{\textcolor{green}{\huge \ding{51}}} & \multicolumn{1}{c|}{\textcolor{green}{\huge \ding{51}}} & \multicolumn{1}{c|}{Function} & heuristic & \multicolumn{1}{c|}{\textcolor{red}{\huge \ding{55}}} & \textcolor{red}{\huge \ding{55}} \\ \hline
Fu et al.~\cite{DBLP:journals/ese/FuTLKNPG24} & <1B & \multicolumn{1}{l|}{} & \multicolumn{1}{l|}{} & \multicolumn{1}{l|}{} & \multicolumn{1}{l|}{} & \multicolumn{1}{l|}{} & \multicolumn{1}{c|}{\textcolor{green}{\huge \ding{51}}} & \multicolumn{1}{l|}{} &  & \multicolumn{1}{c|}{Function} & heuristic & \multicolumn{1}{c|}{\textcolor{red}{\huge \ding{55}}} & \textcolor{red}{\huge \ding{55}} \\ \hline
Sejfia et al.~\cite{DBLP:conf/icse/SejfiaDSM24} & <1B & \multicolumn{1}{l|}{} & \multicolumn{1}{l|}{} & \multicolumn{1}{l|}{} & \multicolumn{1}{l|}{} & \multicolumn{1}{l|}{} & \multicolumn{1}{c|}{\textcolor{green}{\huge \ding{51}}} & \multicolumn{1}{l|}{} &  & \multicolumn{1}{c|}{Function} & \begin{tabular}[c]{@{}c@{}}heuristic, \\partially \\ manually labeled\end{tabular} & \multicolumn{1}{c|}{\textcolor{red}{\huge \ding{55}}} & \textcolor{red}{\huge \ding{55}} \\ \hline
Wang et al.~\cite{wang2023combining} & <1B & \multicolumn{1}{l|}{} & \multicolumn{1}{c|}{\textcolor{green}{\huge \ding{51}}} & \multicolumn{1}{l|}{} & \multicolumn{1}{c|}{\textcolor{green}{\huge \ding{51}}} & \multicolumn{1}{l|}{} & \multicolumn{1}{l|}{} & \multicolumn{1}{l|}{} &  & \multicolumn{1}{c|}{Function} & \begin{tabular}[c]{@{}c@{}}synthetic, \\ heuristic,\\ manually labeled\end{tabular} & \multicolumn{1}{c|}{\textcolor{red}{\huge \ding{55}}} & \textcolor{red}{\huge\ding{55}} \\ \hline
 Liu et al.~\cite{liu2024pre} & <1B & \multicolumn{1}{l|}{} & \multicolumn{1}{c|}{\textcolor{green}{\huge\ding{51}}} & \multicolumn{1}{l|}{} & \multicolumn{1}{c|}{\textcolor{green}{\huge \ding{51}}} & \multicolumn{1}{l|}{} & \multicolumn{1}{l|}{} & \multicolumn{1}{l|}{} &  & \multicolumn{1}{c|}{Function} & \begin{tabular}[c]{@{}c@{}}heuristic,\\ partially \\ manually labeled\end{tabular} & \multicolumn{1}{c|}{\textcolor{red}{\huge \ding{55}}} & \textcolor{red}{\huge \ding{55}} \\ \hline
Rahman et al.~\cite{mahbubur2023towards} & <1B & \multicolumn{1}{l|}{} & \multicolumn{1}{l|}{} & \multicolumn{1}{l|}{} & \multicolumn{1}{l|}{} & \multicolumn{1}{c|}{\textcolor{green}{\huge \ding{51}}} & \multicolumn{1}{l|}{} & \multicolumn{1}{l|}{} &  & \multicolumn{1}{c|}{Function} & \begin{tabular}[c]{@{}c@{}}heuristic,\\ partially \\ manually labeled\end{tabular} & \multicolumn{1}{c|}{\textcolor{red}{\huge \ding{55}}} & \textcolor{red}{\huge \ding{55}} \\ \hline
Ding et al.~\cite{DBLP:journals/corr/abs-2403-18624} & \begin{tabular}[c]{@{}c@{}}<1B,\\ 1-10B,\\ >10B\end{tabular} & \multicolumn{1}{c|}{\textcolor{green}{\huge \ding{51}}} & \multicolumn{1}{l|}{} & \multicolumn{1}{l|}{} & \multicolumn{1}{c|}{\textcolor{green}{\huge \ding{51}}} & \multicolumn{1}{l|}{} & \multicolumn{1}{l|}{} & \multicolumn{1}{l|}{} &  & \multicolumn{1}{c|}{Function} & \begin{tabular}[c]{@{}c@{}}heuristic,\\ partially \\ manually labeled\end{tabular} & \multicolumn{1}{c|}{\textcolor{red}{\huge \ding{55}}} & \textcolor{red}{\huge \ding{55}} \\ \hline
Steenhoek et al.~\cite{DBLP:journals/corr/abs-2311-04109} & <1B & \multicolumn{1}{l|}{} & \multicolumn{1}{l|}{} & \multicolumn{1}{l|}{} & \multicolumn{1}{c|}{\textcolor{green}{\huge \ding{51}}} & \multicolumn{1}{l|}{} & \multicolumn{1}{l|}{} & \multicolumn{1}{l|}{} &  & \multicolumn{1}{c|}{\begin{tabular}[c]{@{}c@{}}Function, \\ Line\end{tabular}} & \begin{tabular}[c]{@{}c@{}}synthetic, \\ heuristic, \\partially \\ manually labeled\end{tabular} & \multicolumn{1}{c|}{\textcolor{red}{\huge \ding{55}}} & \textcolor{red}{\huge \ding{55}} \\ \hline
Khare et al.~\cite{DBLP:journals/corr/abs-2311-16169} & >1B & \multicolumn{1}{l|}{} & \multicolumn{1}{l|}{} & \multicolumn{1}{l|}{} & \multicolumn{1}{l|}{} & \multicolumn{1}{l|}{} & \multicolumn{1}{l|}{} & \multicolumn{1}{c|}{\textcolor{green}{\huge \ding{51}}} &  & \multicolumn{1}{c|}{Function} & \begin{tabular}[c]{@{}c@{}}heuristic, \\partially \\ manually labeled\end{tabular} & \multicolumn{1}{c|}{\textcolor{red}{\huge \ding{55}}} & \textcolor{red}{\huge \ding{55}} \\ \hline
Zhang et al.~\cite{DBLP:journals/corr/abs-2308-12697} & >10B & \multicolumn{1}{l|}{} & \multicolumn{1}{l|}{} & \multicolumn{1}{l|}{} & \multicolumn{1}{l|}{} & \multicolumn{1}{l|}{} & \multicolumn{1}{l|}{} & \multicolumn{1}{c|}{\textcolor{green}{\huge \ding{51}}} &  & \multicolumn{1}{c|}{Function} & \begin{tabular}[c]{@{}c@{}}synthetic, \\heuristic\end{tabular} & \multicolumn{1}{c|}{\textcolor{red}{\huge \ding{55}}} & \textcolor{red}{\huge \ding{55}} \\ \hline
Risse et al.~\cite{DBLP:conf/uss/RisseB24} & <1B & \multicolumn{1}{l|}{} & \multicolumn{1}{l|}{} & \multicolumn{1}{l|}{} & \multicolumn{1}{l|}{} & \multicolumn{1}{l|}{} & \multicolumn{1}{c|}{\textcolor{green}{\huge \ding{51}}} & \multicolumn{1}{l|}{} &  & \multicolumn{1}{c|}{Function} & \begin{tabular}[c]{@{}c@{}}partially manually \\ labeled,\\ manually labeled\end{tabular} & \multicolumn{1}{c|}{\textcolor{red}{\huge \ding{55}}} & \textcolor{red}{\huge \ding{55}} \\ \hline
Steenhoek et al.~\cite{DBLP:conf/icse/SteenhoekRJL23} & <1B & \multicolumn{1}{l|}{} & \multicolumn{1}{l|}{} & \multicolumn{1}{l|}{} & \multicolumn{1}{l|}{} & \multicolumn{1}{l|}{} & \multicolumn{1}{c|}{\textcolor{green}{\huge \ding{51}}} & \multicolumn{1}{l|}{} &  & \multicolumn{1}{c|}{\begin{tabular}[c]{@{}c@{}}Function,\\ Line\end{tabular}} & \begin{tabular}[c]{@{}c@{}}heuristic \\, partially \\ manually labeled\end{tabular} & \multicolumn{1}{c|}{\textcolor{red}{\huge \ding{55}}} & \textcolor{red}{\huge \ding{55}} \\ \hline
Yang et al.~\cite{DBLP:conf/icse/YangWLW23} & <1B & \multicolumn{1}{c|}{\textcolor{green}{\huge \ding{51}}} & \multicolumn{1}{l|}{} & \multicolumn{1}{l|}{} & \multicolumn{1}{l|}{} & \multicolumn{1}{l|}{} & \multicolumn{1}{l|}{} & \multicolumn{1}{l|}{} &  & \multicolumn{1}{c|}{Function} & \begin{tabular}[c]{@{}c@{}}heuristic,\\partially \\ manually labeled\end{tabular} & \multicolumn{1}{c|}{\textcolor{red}{\huge \ding{55}}} & \textcolor{red}{\huge \ding{55}} \\ \hline
Chen et al.~\cite{DBLP:conf/raid/0001DACW23} & <1B & \multicolumn{1}{l|}{} & \multicolumn{1}{l|}{} & \multicolumn{1}{l|}{} & \multicolumn{1}{l|}{} & \multicolumn{1}{l|}{} & \multicolumn{1}{c|}{\textcolor{green}{\huge \ding{51}}} & \multicolumn{1}{l|}{} &  & \multicolumn{1}{c|}{Function} & heuristic & \multicolumn{1}{c|}{\textcolor{red}{\huge \ding{55}}} & \textcolor{red}{\huge \ding{55}} \\ \hline
Li et al.~\cite{DBLP:conf/icse/0027WZLZX0024} & <1B & \multicolumn{1}{l|}{} & \multicolumn{1}{l|}{} & \multicolumn{1}{l|}{} & \multicolumn{1}{l|}{} & \multicolumn{1}{l|}{} & \multicolumn{1}{c|}{\textcolor{green}{\huge \ding{51}}} & \multicolumn{1}{l|}{} &  & \multicolumn{1}{c|}{Function} & heuristic & \multicolumn{1}{c|}{\textcolor{red}{\huge \ding{55}}} & \textcolor{red}{\huge \ding{55}} \\ \hline
Croft et al.~\cite{DBLP:conf/icse/CroftBK23} & <1B & \multicolumn{1}{l|}{} & \multicolumn{1}{l|}{} & \multicolumn{1}{l|}{} & \multicolumn{1}{l|}{} & \multicolumn{1}{l|}{} & \multicolumn{1}{c|}{\textcolor{green}{\huge \ding{51}}} & \multicolumn{1}{l|}{} &  & \multicolumn{1}{c|}{Function} &  \begin{tabular}[c]{@{}c@{}}heuristic,\\ manually labeled \\\end{tabular} & \multicolumn{1}{c|}{\textcolor{red}{\huge\ding{55}}} & \textcolor{red}{\huge \ding{55}} \\ \hline 
Le et al.~\cite{DBLP:conf/msr/LeDB24} & <1B & \multicolumn{1}{l|}{} & \multicolumn{1}{l|}{} & \multicolumn{1}{l|}{} & \multicolumn{1}{l|}{} & \multicolumn{1}{l|}{} & \multicolumn{1}{c|}{\textcolor{green}{\huge \ding{51}}} & \multicolumn{1}{l|}{} &  & \multicolumn{1}{c|}{\begin{tabular}[c]{@{}c@{}}Function, \\ Line\end{tabular}} &  \begin{tabular}[c]{@{}c@{}}heuristic,\\ partially \\ manually labeled\end{tabular} & \multicolumn{1}{c|}{\textcolor{red}{\huge \ding{55}}} & \textcolor{red}{\huge \ding{55}} \\ \hline
Zhou et al.~\cite{zhou2024comparison} & \begin{tabular}[c]{@{}c@{}}<10B\end{tabular} & \multicolumn{1}{l|}{} & \multicolumn{1}{l|}{} & \multicolumn{1}{l|}{} & \multicolumn{1}{l|}{} & \multicolumn{1}{l|}{} & \multicolumn{1}{c|}{\textcolor{green}{\huge \ding{51}}} & \multicolumn{1}{c|}{\textcolor{green}{\huge \ding{51}}} &  & \multicolumn{1}{c|}{Repo} &  \begin{tabular}[c]{@{}c@{}} heuristic,\\ manually labeled \\\end{tabular} & \multicolumn{1}{c|}{\textcolor{red}{\huge \ding{55}}} & \textcolor{red}{\huge \ding{55}} \\ \hline
Tran et al.~\cite{tran2024detectvul} & <1B & \multicolumn{1}{l|}{} & \multicolumn{1}{c|}{\textcolor{green}{\huge\ding{51}}} & \multicolumn{1}{l|}{} & \multicolumn{1}{l|}{} & \multicolumn{1}{l|}{} & \multicolumn{1}{c|}{} & \multicolumn{1}{c|}{} &  & \multicolumn{1}{c|}{Line} &  heuristic & \multicolumn{1}{c|}{\textcolor{red}{\huge \ding{55}}} & \textcolor{red}{\huge \ding{55}} \\ \hline
Jiang et al~\cite{DBLP:conf/internetware/JiangSGWW0024} & <1B & \multicolumn{1}{l|}{} & \multicolumn{1}{c|}{} & \multicolumn{1}{c|}{\textcolor{green}{\huge \ding{51}}} & \multicolumn{1}{l|}{} & \multicolumn{1}{l|}{} & \multicolumn{1}{c|}{} & \multicolumn{1}{c|}{} &  & \multicolumn{1}{c|}{Function} &  \begin{tabular}[c]{@{}c@{}}heuristic,\\ partially \\ manually labeled\end{tabular} & \multicolumn{1}{c|}{\textcolor{red}{\huge \ding{55}}} & \textcolor{red}{\huge \ding{55}} \\ \hline
Shestov et al.~\cite{DBLP:journals/corr/abs-2401-17010} & >10B & \multicolumn{1}{l|}{} & \multicolumn{1}{c|}{} & \multicolumn{1}{l|}{} & \multicolumn{1}{l|}{} & \multicolumn{1}{c|}{} & \multicolumn{1}{c|}{\textcolor{green}{\huge \ding{51}}} & \multicolumn{1}{c|}{} & \multicolumn{1}{c|}{} & \multicolumn{1}{c|}{Function} &  \begin{tabular}[c]{@{}c@{}}heuristic,\\ manually labeled\end{tabular} & \multicolumn{1}{c|}{\textcolor{red}{\huge \ding{55}}} & \textcolor{red}{\huge \ding{55}} \\ \hline
Ni et al~\cite{ni2024learning} & >10B & \multicolumn{1}{l|}{} & \multicolumn{1}{c|}{} & \multicolumn{1}{l|}{} & \multicolumn{1}{l|}{} & \multicolumn{1}{c|}{} & \multicolumn{1}{c|}{} & \multicolumn{1}{c|}{\textcolor{green}{\huge \ding{51}}} & \multicolumn{1}{c|}{} & \multicolumn{1}{c|}{Function} &  \begin{tabular}[c]{@{}c@{}}heuristic\end{tabular} & \multicolumn{1}{c|}{\textcolor{red}{\huge \ding{55}}} & \textcolor{red}{\huge \ding{55}} \\ \hline
Weng et al.~\cite{DBLP:conf/internetware/WengQLLC24} & <1B & \multicolumn{1}{l|}{} & \multicolumn{1}{c|}{\textcolor{green}{\huge \ding{51}}} & \multicolumn{1}{l|}{} & \multicolumn{1}{l|}{} & \multicolumn{1}{c|}{} & \multicolumn{1}{c|}{} & \multicolumn{1}{c|}{} & \multicolumn{1}{c|}{} & \multicolumn{1}{c|}{Line} &  \begin{tabular}[c]{@{}c@{}}heuristic\end{tabular} & \multicolumn{1}{c|}{\textcolor{red}{\huge \ding{55}}} & \textcolor{red}{\huge \ding{55}} \\ \hline
Du et al.~\cite{DBLP:journals/corr/abs-2406-11147} & >10B & \multicolumn{1}{l|}{} & \multicolumn{1}{c|}{} & \multicolumn{1}{c|}{} & \multicolumn{1}{l|}{} & \multicolumn{1}{c|}{} & \multicolumn{1}{c|}{} & \multicolumn{1}{c|}{} & \multicolumn{1}{c|}{\textcolor{green}{\huge \ding{51}}} & \multicolumn{1}{c|}{Function} &  \begin{tabular}[c]{@{}c@{}}heuristic\end{tabular} & \multicolumn{1}{c|}{\textcolor{red}{\huge \ding{55}}} & \textcolor{red}{\huge \ding{55}} \\ \hline
Wen et al.~\cite{DBLP:journals/corr/abs-2404-15596} & \begin{tabular}[c]{@{}c@{}}<1B,\\ 1-10B, \\ >10B\end{tabular} & \multicolumn{1}{l|}{} & \multicolumn{1}{c|}{} & \multicolumn{1}{l|}{} & \multicolumn{1}{l|}{} & \multicolumn{1}{c|}{} & \multicolumn{1}{c|}{} & \multicolumn{1}{c|}{} & \multicolumn{1}{c|}{\textcolor{green}{\huge \ding{51}}} & \multicolumn{1}{c|}{Function} &  \begin{tabular}[c]{@{}c@{}}heuristic\end{tabular} & \multicolumn{1}{c|}{\textcolor{red}{\huge \ding{55}}} & \textcolor{red}{\huge \ding{55}} \\ \hline
Xin Yin~\cite{DBLP:journals/corr/abs-2404-03994} & \begin{tabular}[c]{@{}c@{}}>10B\end{tabular} & \multicolumn{1}{l|}{} & \multicolumn{1}{c|}{} & \multicolumn{1}{l|}{} & \multicolumn{1}{l|}{} & \multicolumn{1}{c|}{} & \multicolumn{1}{c|}{} & \multicolumn{1}{c|}{\textcolor{green}{\huge \ding{51}}} & \multicolumn{1}{c|}{} & \multicolumn{1}{c|}{Function} &  \begin{tabular}[c]{@{}c@{}}heuristic\end{tabular} & \multicolumn{1}{c|}{\textcolor{red}{\huge \ding{55}}} & \textcolor{red}{\huge \ding{55}} \\ \hline
Yang et al.~\cite{yang2024security} & \begin{tabular}[c]{@{}c@{}}<1B,\\>10B\end{tabular} & \multicolumn{1}{l|}{} & \multicolumn{1}{c|}{} & \multicolumn{1}{c|}{\textcolor{green}{\huge \ding{51}}} & \multicolumn{1}{l|}{} & \multicolumn{1}{c|}{} & \multicolumn{1}{c|}{} & \multicolumn{1}{c|}{} & \multicolumn{1}{c|}{} & \multicolumn{1}{c|}{Function} &  \begin{tabular}[c]{@{}c@{}}heuristic\end{tabular} & \multicolumn{1}{c|}{\textcolor{green}{\huge \ding{51}}} & \textcolor{red}{\huge \ding{55}} \\  
\bottomrule
\end{tabular}
}
\label{table4detection}
\end{table*}

In the following subsection, we will introduce detailed categories of LLM adaptation techniques for each of the following: \emph{1) fine-tuning}, \emph{2) prompt engineering}, and \emph{3) retrieval augmentation (RAG)}.
Table~\ref{table4detection} summarizes the existing LLM-based methods for vulnerability detection.

\subsection{\textbf{Fine-tuning}}
\label{sec_detection:fine-tuning}
Fine-tuning is a commonly used technique for adapting LLMs to vulnerability detection. In this process, labeled code samples, indicating whether they are vulnerable or not, are provided as training data. The model is then fine-tuned through supervised learning, where its parameters are adjusted based on these labeled examples. Several vulnerability detection studies~\cite{DBLP:conf/msr/FuT22,DBLP:conf/acsac/ThapaJACPN22,DBLP:journals/ese/FuTLKNPG24,DBLP:conf/icse/SejfiaDSM24,DBLP:conf/uss/RisseB24,DBLP:conf/icse/SteenhoekRJL23,DBLP:conf/raid/0001DACW23,DBLP:conf/icse/0027WZLZX0024,DBLP:conf/icse/CroftBK23, DBLP:conf/msr/LeDB24,zhou2024comparison,DBLP:journals/corr/abs-2401-17010} have directly applied this default fine-tuning approach without introducing additional designs. In this subsection, we focus on studies that propose advanced fine-tuning techniques that go beyond the default approach for improved effectiveness.

The fine-tuning process usually involves several steps/stages, such as data preparation, model design, model training, and model evaluation.
Regarding fine-tuning, we classify adaptation techniques into five groups based on the stages they mainly target:
\emph{Data-centric innovations} (data preparation), \emph{Combination with program analysis} (data preparation), \emph{LLM+ other deep learning modules} (model design),  \emph{Domain-specific pre-training} (model training), and \emph{Causal learning} (training optimization).

\vspace{0.2cm}
\noindent\textbf{Data-centric Innovations.} 
Data-centric innovations focus on optimizing the vulnerability detection data used for fine-tuning LLMs. Prior studies~\cite{DBLP:conf/icse/YangWLW23, DBLP:conf/icse/CroftBK23,DBLP:conf/issta/NieLWWLW23,DBLP:conf/raid/0001DACW23,DBLP:conf/scam/KuangYZTY23} have found that existing vulnerability detection data can suffer from imbalanced label distribution, noisy or incorrect labels, and scarcity of labeled data. Researchers~\cite{DBLP:conf/icse/YangWLW23,DBLP:conf/kbse/WenWGWLG23,DBLP:conf/scam/KuangYZTY23,DBLP:journals/corr/abs-2403-18624} have explored how to address the data issues.
\vspace{-\topsep}
\begin{itemize}[leftmargin=*]

\item \textit{Imbalanced Learning}: To address label imbalance, i.e., having more non-vulnerable code samples than vulnerable ones in the dataset, Yang et al.~\cite{DBLP:conf/icse/YangWLW23} applied various data sampling techniques. In addition, Ding et al.~\cite{DBLP:journals/corr/abs-2403-18624} up-weighted the loss value for the rare class (i.e., the vulnerable samples) to pay comparable attention to both vulnerable and clean classes.
They found that random oversampling on raw code data enhances the ability of the LLM-based vulnerability detection approach to learn real vulnerable patterns.

\item \textit{Positive and Unlabeled Learning}: To address label quality issues such as noisy or incorrect labels, Wen et al.~\cite{DBLP:conf/kbse/WenWGWLG23} proposed PILOT, which learns solely from positive (vulnerable) and unlabeled data for vulnerability detection. Specifically, PILOT generates pseudo-labels for selected unlabeled data and mitigates the data noise by using a mixed-supervision loss.

\item \textit{Counterfactual Training}: To enhance the diversity of labeled data, Kuang et al.~\cite{DBLP:conf/scam/KuangYZTY23} proposed perturbing user-defined identifiers in the source code while preserving the syntactic and semantic structure. This approach generates diverse counterfactual training data, which refers to hypothetical data (e.g., data after perturbing identifiers) differing from actual data (i.e., data without perturbation), useful for analyzing the effect of certain factors.
Incorporating these counterfactual data enriches the training data for LLMs.
\end{itemize}
\vspace{-\topsep}

\vspace{0.2cm}
\noindent\textbf{Combination of LLM with Program Analysis.}
Many LLMs undergo pre-training on extensive datasets through unsupervised objectives like masked language modeling~\cite{devlin2018bert} or next token prediction~\cite{radford2019language}. For those LLMs, they may prioritize capturing sequential features and could potentially overlook certain structural aspects crucial for understanding code. To address this limitation, several studies have proposed integrating program analysis techniques with LLMs. The idea involves utilizing program analysis to extract structural features/relations within code, which are then incorporated into LLMs to enhance their understanding.
Specifically, Liu et al.~\cite{liu2024pre} leveraged Joern~\cite{Joern} to build the AST and PDG of the function, leveraging this data to pre-train their LLM to predict statement-level control dependencies and token-level data dependencies within the function. 
Peng et al.~\cite{DBLP:conf/scam/PengCZTLH23} utilized program slicing to extract control and data dependency information, aiding LLMs in vulnerability detection. 
Wang et al.~\cite{wang2023combining} proposed to learn program presentations by feeding static source code information and dynamic program execution traces with LLMs.
Additionally, Zhang et al.~\cite{DBLP:journals/tse/ZhangLHXL23} proposed decomposing syntax-based Control Flow Graphs (CFGs) into multiple execution paths and feeding these paths to LLMs for vulnerability detection.
Tran et al.~\cite{tran2024detectvul} utilize a program analysis tool to normalize all user-defined names, such as variables and functions, into symbolic names (e.g., VAR\_1, VAR\_2, and FUNC\_1) to enhance the model's robustness against adversarial attacks.
Weng et al.~\cite{DBLP:conf/internetware/WengQLLC24} incorporated inter-statement data and control dependency information into an LLM by masking irrelevant attention scores based on the program dependency graph, which resulted in improved code representation.

\vspace{0.2cm}
\noindent\textbf{Combination of LLM with Other Deep Learning Modules.}
LLMs have their own inherent limitations. Firstly, most LLMs are based on the Transformer architecture, which primarily models sequential relations and features. 
Secondly, some LLMs (e.g., CodeBERT) impose restrictions on the length of input code snippets. For instance, the most frequently used LLM in vulnerability detection, CodeBERT, can only process 512 tokens.
To address these limitations, researchers have attempted to combine LLMs with other DL modules: 
\vspace{-\topsep}
\begin{itemize} [leftmargin=*]
\item \textit{LLM+GNN}:  To leverage the structural features of code more effectively, Tang et al.~\cite{DBLP:journals/jss/TangTBZF23} introduced CSGVD, which employs graph neural networks (GNN) to extract the graph features of code and combine them with the features extracted by CodeBERT. 
Jiang et al.~\cite{DBLP:conf/internetware/JiangSGWW0024} parsed data flow graphs (DFGs) from the input code and used the data types of each variable as the node characteristics of the DFG. They then embedded the DFG information into the graph representation using a graph neural network (GNN) and integrated this graph representation with a large language model.
In addition, Yang et al.~\cite{yang2024security} combined an LLM with a GNN model. Specifically, they first used data flow analysis (DFA) embeddings to establish a GNN model. They then concatenated the learned embeddings from the GNN with the hidden states of the LLM. By concatenating the embeddings during each forward pass, they enabled simultaneous training of both the LLM and the GNN.

\item \textit{LLM+Bi-LSTM}: To address the length constraints of LLMs, Ziems et al.~\cite{DBLP:conf/infocom/ZiemsW21} first segmented the input code into multiple fixed-size segments. They then utilized BERT to encode each segment and incorporated a Bidirectional Long Short-Term Memory (Bi-LSTM) module to process the output of BERT on each segment. Finally, a softmax classifier was applied to the last hidden state of the Bi-LSTM to produce the final classification scores.
\end{itemize}
\vspace{-\topsep}

\vspace{0.2cm}
\noindent\textbf{Domain-specific Pre-training.}
Domain-specific pre-training involves pre-training an LLM on data specific to a particular domain, such as vulnerability data, before fine-tuning it for a specific task within that domain. This process enables the LLM to better understand the data relevant to the domain.
Several studies in the field of vulnerability detection have adopted this technique~\cite{DBLP:conf/ijcnn/HanifM22,liu2024pre,ni2023distinguishing,wang2023combining,DBLP:journals/corr/abs-2403-18624,DBLP:journals/corr/abs-2311-04109}.
These studies typically use one of three pre-training objectives:
\vspace{-\topsep}
\begin{itemize}[leftmargin=*]
    \item \textit{Masked Language Modeling}: This pre-training objective involves training the LLM to predict masked tokens in a corrupted code snippet. Hanif and Mahmood~\cite{DBLP:conf/ijcnn/HanifM22} introduced VulBERTa, which pre-trained a RoBERTa~\cite{liu2019roberta} model on open-source C/C++ projects using the Masked Language Modeling objective. Since C/C++ is the programming language of the vulnerability detection data used to evaluate VulBERTa, this pre-training enhances VulBERTa's ability to understand C/C++ code.
    
    \item \textit{Contrastive Learning:} 
    This pre-training objective is to minimize the distance between similar functions while maximizing the distance between dissimilar functions. Through contrastive learning, the LLM can learn to capture the distinctive features of code. 
    In particular, Ni et al.~\cite{ni2023distinguishing} and Wang et al.~\cite{wang2023combining} utilized different hidden dropout masks to transform the same input function into positive samples (i.e., those similar to the input function) while regarding other distinct functions as dissimilar samples.
    Then they pre-trained an LLM following the contrastive learning objective.
    Additionally, Ding et al.~\cite{DBLP:journals/corr/abs-2403-18624} performed Class-aware Contrastive Learning (CA-CLR) to maximize the representation similarity between each sample and the perturbed version of itself and minimize the representation similarity between two randomly chosen samples.

    \item \textit{Predicting Program Dependencies:} This pre-training objective aims to enhance the LLM by guiding the model to learn the knowledge necessary for analyzing dependencies in programs. Specifically, Liu et al.~\cite{liu2024pre} pre-trained an encoder-only LLM on code from open-source C/C++ projects to predict the statement-level control dependency and token-level data dependency.

    \item \textit{Annotating Vulnerable Statements:} This pre-training objective assumes that improving alignment to potentially vulnerable statements (PVS) would improve the performance of the model. For example, Steenhoek et al.~\cite{DBLP:journals/corr/abs-2311-04109} developed two annotation approaches (i.e., Mark and Prepend) to give guidance or extra information to the model. Mark inserted special ``marker'' tokens before and after each token inside the PVS, while Prepend added the tokens inside PVS at the beginning of the code. This can be considered as inserting domain knowledge into the model’s input.

\end{itemize}
\vspace{-\topsep}
After the pre-training, those domain-specific pre-trained LLMs undergo fine-tuning on the vulnerability detection dataset to perform vulnerability detection.

\vspace{0.2cm}
\noindent\textbf{Causal Learning.} Although promising, LLMs have been found to lack robustness under perturbation or when encountering out-of-distribution (OOD) data~\cite{mahbubur2023towards}. Rahman et al.~\cite{mahbubur2023towards} suggested that this weak robustness may be attributed to LLMs learning non-robust features, such as variable names, that have spurious correlations with labels. To address this issue, Rahman et al. proposed CausalVul, which first designs perturbations to identify spurious features and then applies causal learning algorithms, specifically do-calculus, on top of LLMs to promote causal-based prediction, enhancing the robustness of LLMs in vulnerability detection.

\subsection{\textbf{Prompt Engineering.}}
\label{sec_detection:PromptEngineering}

\vspace{0.2cm}
\noindent\textbf{Zero-shot Prompting.}
Researchers have attempted to devise effective prompts to guide LLMs in conducting vulnerability detection in the zero-shot prompting manner. Their prompt engineering designs generally consist of one or more components outlined below:

\begin{itemize}
    \item \textit{Task Description.} This part of the prompt aims to equip LLMs with valuable task-specific information related to vulnerability detection, clarifying the objectives the model should achieve (e.g., identifying vulnerabilities in the provided code). Different studies employ varying task descriptions, and there is no consensus on which approach is optimal. 
    Specifically, Zhou et al.~\cite{DBLP:journals/corr/abs-2401-15468} used a task description: \textit{``If has any potential vulnerability, output: `this code is vulnerable’. Otherwise, output: `this code is non-vulnerable’. The code is [code]. Let’s start:''}. Fu et al.~\cite{fu2023chatgpt} proposed a task description: \textit{``Predict Whether the C/C++ function below is vulnerable. Strictly return 1 for a vulnerable function and 0 for a non-vulnerable function without further explanation.''}  Purba et al.~\cite{DBLP:conf/issre/PurbaGRC23} devised a task description: \textit{``[code] Is this code vulnerable? Answer in only Yes or No''}. \textit{``[code]''} refers to the input code in these task descriptions. Xin Yin~\cite{DBLP:journals/corr/abs-2404-03994} used the task description: \textit{``I will provide you a C code snippet and want you to tell whether it has a vulnerability. You need to output “yes” or “no” first (output no if uncertain), and then explain.''}
    Zhou et al.~\cite{zhou2024comparison} used \textit{``If the following code snippet has any vulnerabilities, output Yes; otherwise, output No''}.

    \item \textit{Role Description.} This aims to help LLMs shift their operational mode from a general language model to that of a vulnerability detector. Specifically, Zhou et al.~\cite{DBLP:journals/corr/abs-2401-15468} defined a role description: \textit{``You are an experienced developer who knows the security vulnerability very well''}.  Fu et al.~\cite{fu2023chatgpt} used a role description: \textit{``I want you to act as a vulnerability detection system''}. Khare et al.~\cite{DBLP:journals/corr/abs-2311-16169} utilized a role description: \textit{ You are a security researcher''}. 
    Xin Yin~\cite{DBLP:journals/corr/abs-2404-03994} used the role description as: \textit{``I want you to act as a vulnerability scanner''}.

     \item \textit{Vulnerability-Related Auxiliary Information.} Some studies have integrated vulnerability-related auxiliary information into their prompt designs to enhance LLM performance. Specifically, Zhou et al.\cite{DBLP:journals/corr/abs-2401-15468} suggested including vulnerable code examples representing the top 25 most dangerous CWEs of 2022 as part of the prompts to help LLMs better understand the characteristics of serious vulnerabilities. Additionally, Khare et al.\cite{DBLP:journals/corr/abs-2311-16169} specified the types of vulnerabilities when querying LLMs about potential vulnerabilities in the provided code.

     \item \textit{Program-Analysis Auxiliary Information.} Some studies utilize program-analysis tools to extract code dependencies from the input code, aiming to enhance LLMs' understanding of the input code. Specifically, Khare et al.~\cite{DBLP:journals/corr/abs-2311-16169} designed a dataflow analysis-based prompt that guides the model to simulate a source-sink-sanitizer-based dataflow analysis on the target code snippet before predicting if it is vulnerable. Zhang et al.~\cite{DBLP:journals/corr/abs-2308-12697} designed prompts that incorporate auxiliary information on data flow or API calls.

     \item \textit{Chain-of-thought (CoT) Prompting.} It is proposed by Kojima et al.~\cite{DBLP:conf/nips/KojimaGRMI22} for improved reasoning, which involves adding ``Let’s think step by step'' to the original prompt. Specifically, Zhou et al.~\cite{zhou2024comparison} and Ni et al.~\cite{ni2024learning} added the ``Let’s think step by step'' into the prompt.

\end{itemize}

\vspace{0.2cm}
\noindent\textbf{Few-shot Prompting.}  In this approach, a few examples of input and ground truth label pairs are provided to the large language models (LLMs) as additional guidance. Specifically, Ni et al.~\cite{ni2024learning} utilize between 1 and 6 few-shot learning examples to fill the fixed context window (i.e., 4,096 tokens). Zhou et al.~\cite{zhou2024comparison} use two few-shot examples.

\subsection{\textbf{Retrieval Augmentation (RAG).}}
\label{sec_detection:RetrievalAugmentation}

Retrieval augmentation is a technique used to enhance the few-shot prompting. It involves retrieving similar labeled data samples from the training set when given a test data sample and using these retrieved data as examples to guide the prediction of LLMs on the test sample.
Liu et al.~\cite{DBLP:conf/dsc/LiuLGG23} proposed using efficient retrieval tools such as BM-25 and TF-IDF for retrieval.
Zhou et al.~\cite{DBLP:journals/corr/abs-2401-15468} suggested using CodeBERT as a retrieval tool. This method first transforms code snippets into semantic vectors and then quantifies the similarity between two code snippets by calculating the Cosine similarity of their respective semantic vectors. It finally returned the top similar code based on the similarity scores.
Du et al.~\cite{DBLP:journals/corr/abs-2406-11147} leverage a knowledge-level retrieval-augmentation framework consisting of three phases to identify vulnerabilities in a given code. First, they constructed a vulnerability knowledge base by extracting multi-dimensional knowledge from existing CVE instances using LLMs. Second, for a given code snippet, they retrieved relevant vulnerability knowledge from the constructed knowledge base based on functional semantics.
Finally, they utilized LLMs to assess the vulnerability of the given code snippet by reasoning about the presence of vulnerability causes and proposing potential fixing solutions derived from the retrieved knowledge.
Wen et al.~\cite{DBLP:journals/corr/abs-2404-15596} first retrieved the most relevant dependencies from call graphs based on a given input code, and then integrated these dependencies with the input code into LLMs to detect vulnerabilities.

\begin{tcolorbox}[left=4pt,right=4pt,top=2pt,bottom=2pt,boxrule=0.5pt]
\textbf{Answer to RQ2:} 
Our analysis revealed three commonly used techniques to adapt LLMs for vulnerability detection: including \emph{fine-tuning} ($\approx$73\%), \emph{prompt engineering} ($\approx$17\%), and \emph{retrieval augmentation} ($\approx$10\%).
\end{tcolorbox}

\begin{figure}[t]
  \centering
  \includegraphics[width=.5\linewidth]{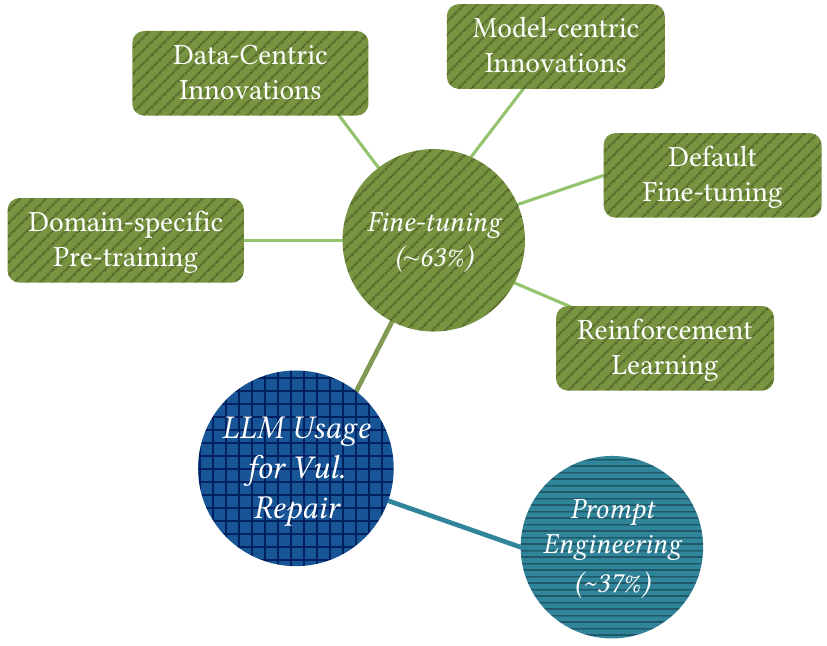}
\caption{Adaptation Techniques of LLMs for Vul. Repair}
\label{ExpTec2}
\end{figure}

\section{RQ3: How Are LLMs Adapted for Vulnerability Repair?}\label{RQ3}

Regarding the usage of LLMs, we summarize \textbf{\emph{three major categories}} from the included studies: \textbf{1) fine-tuning} (approximately 63\%), which updates the parameters of LLMs using a labeled dataset, and \textbf{2) prompt engineering} (approximately 37\%), which freezes the parameters of LLMs and designs prompts to guide LLMs in generating relevant responses.

In the following subsection, we will introduce detailed categories of LLM adaptation techniques for each of the following: \emph{1) fine-tuning} and \emph{2) prompt engineering}.

\subsection{\textbf{Fine-tuning}}
\label{sec_repair:fine-tuning}

Some studies~\cite{DBLP:conf/sigsoft/FuTLNP22,DBLP:journals/ese/FuTLKNPG24} employ default fine-tuning methods to adapt LLMs for vulnerability repair. In this subsection, we focus on research that introduces advanced fine-tuning techniques that extend beyond the default approach to achieve improved effectiveness.

Regarding fine-tuning, we categorize advanced adaptation techniques into four groups based on the stages they mainly target:
\emph{Data-centric innovations} (data preparation), \emph{Model-centric innovations} (model design),  \emph{Domain-specific pre-training} (model training), and \emph{Reinforcement learning} (training optimization).
Table~\ref{table4repair} summarizes the existing LLM-based methods for vulnerability repair.

\vspace{0.2cm}
\noindent\textbf{Data-Centric Innovations.}
Several studies have shown that in addition to the input vulnerable code, incorporating other diverse relevant inputs can boost the effectiveness of LLMs.
Specifically, the AST of the vulnerable code input~\cite{zhou2022spvf}, vulnerability descriptions~\cite{zhou2022spvf,DBLP:journals/ejwcn/WeiBWLYSL23}, and vulnerable code examples shared on CWE websites~\cite{zhou2024out} can enhance the LLMs in repairing vulnerability.
In addition, Wei et al.~\cite{DBLP:journals/ejwcn/WeiBWLYSL23} found that vulnerability-inducing commits and vulnerability-fixing commits can also improve the effectiveness of the model.
Lastly, Zhou et al.~\cite {zhou2024out} found that certain vulnerable functions exceed the length limit of Transformer-based LLMs like CodeT5 (512 subtokens)~\cite{zhou2024out}. To address this limitation, they applied the Fusion-in-Decoder framework to partition a long code function into multiple segments and feed those segments into LLMs one by one. Their findings demonstrated that those segments can boost model effectiveness.
Additionally, a study shows that removing less important parts of the code can help the model focus more effectively on the vulnerable lines, thereby improving performance.
Berabi et al~\cite{DBLP:journals/corr/abs-2402-13291} presented a method for generating reduced code by leveraging static analysis outcomes, creating a compact representation of the program. This smaller version fits within the attention window of an LLM while preserving the crucial information required to learn a correct fix. The technique ensures that the reduced code snippet still produces the same static analysis report as the original program, maintaining the core problem for effective analysis.

\vspace{0.2cm}
\noindent\textbf{Model-Centric Innovations.}
Model-centric innovations encompass methodologies that prioritize revising the model architecture of LLMs (i.e., the Transformer~\cite{vaswani2017attention}). Specifically, Fu et al.~\cite{fu2023vision} drew inspiration from Vision Transformer (VIT)-based object detection techniques in computer vision. They proposed vulnerability queries within the pre-trained Transformer model to identify vulnerable code blocks within the source code. Additionally, they trained a model to learn a vulnerability mask, enhancing the attention of the vulnerability queries on the vulnerable code areas.
Islam et al.~\cite{DBLP:journals/corr/abs-2401-03374} proposed fine-tuning an LLM (CodeGen2~\cite{DBLP:journals/corr/abs-2305-02309}) to perform two tasks simultaneously. First, the model generates a fix for the vulnerable code, and second, it produces a developer-friendly description of the code as an explanation.

\begin{table*}[t]
\caption{Existing LLM-based Methods for Vulnerability Repair}
\resizebox{0.55\width}{!}{%
\begin{tabular}{l|c|cccccc|cc|cc}
\toprule
 & \textbf{LLMs} & \multicolumn{6}{c|}{\textbf{LLM Adaptation Techniques}} & \multicolumn{2}{c|}{\textbf{Data}} & \multicolumn{2}{c}{\textbf{Deployment}} \\ \hline
 & \multicolumn{1}{c|}{\textbf{Size}} & \multicolumn{1}{c|}{\begin{tabular}[c]{@{}c@{}}\textbf{Data-}\\ \textbf{centric}\end{tabular}} & \multicolumn{1}{c|}{\begin{tabular}[c]{@{}c@{}}\textbf{Model-}\\ \textbf{centric}\end{tabular}} & \multicolumn{1}{c|}{\begin{tabular}[c]{@{}c@{}}\textbf{Domain}\\ \textbf{Pre-training}\end{tabular}} & \multicolumn{1}{c|}{\begin{tabular}[c]{@{}c@{}}\textbf{Reinforce.}\\ \textbf{Learning}\end{tabular}} & \multicolumn{1}{c|}{\begin{tabular}[c]{@{}c@{}}\textbf{Default} \\ \textbf{Tuning}\end{tabular}} & \begin{tabular}[c]{@{}c@{}}\textbf{Prompt} \\ \textbf{Eng.}\end{tabular} & \multicolumn{1}{c|}{\begin{tabular}[c]{@{}c@{}}\textbf{Input} \\ \textbf{Granularity}\end{tabular}} & \begin{tabular}[c]{@{}c@{}}\textbf{Real-world\&}\\ \textbf{Tests}\end{tabular} & \multicolumn{1}{c|}{\begin{tabular}[c]{@{}c@{}}\textbf{Interact}\\\textbf{with Users}\end{tabular}} & \begin{tabular}[c]{@{}c@{}}\textbf{Integrated}\\ \textbf{to Workflow}\end{tabular} \\ \hline
Chen et al.~\cite{DBLP:journals/tse/ChenKM23} & \textless{}1B & \multicolumn{1}{c|}{} & \multicolumn{1}{c|}{} & \multicolumn{1}{c|}{\textcolor{green}{\huge \ding{51}}} & \multicolumn{1}{c|}{} & \multicolumn{1}{c|}{} &  & \multicolumn{1}{c|}{Function} & \textcolor{red}{\huge \ding{55}} & \multicolumn{1}{c|}{\textcolor{red}{\huge \ding{55}}} & \textcolor{red}{\huge \ding{55}} \\ \hline
Chi et al.~\cite{DBLP:journals/tse/ChiQLZY23} & \textless{}1B & \multicolumn{1}{c|}{} & \multicolumn{1}{c|}{} & \multicolumn{1}{c|}{\textcolor{green}{\huge \ding{51}}} & \multicolumn{1}{c|}{} & \multicolumn{1}{c|}{} &  & \multicolumn{1}{c|}{Function} & \textcolor{red}{\huge \ding{55}} & \multicolumn{1}{c|}{\textcolor{red}{\huge \ding{55}}} & \textcolor{red}{\huge \ding{55}} \\ \hline
Fu et al.~\cite{DBLP:conf/sigsoft/FuTLNP22} & \textless{}1B & \multicolumn{1}{c|}{} & \multicolumn{1}{c|}{} & \multicolumn{1}{c|}{} & \multicolumn{1}{c|}{} & \multicolumn{1}{c|}{\textcolor{green}{\huge \ding{51}}} &  & \multicolumn{1}{c|}{Function} & \textcolor{red}{\huge \ding{55}} & \multicolumn{1}{c|}{\textcolor{red}{\huge \ding{55}}} & \textcolor{red}{\huge \ding{55}} \\ \hline
Zhou et al.~\cite{zhou2022spvf} & \textless{}1B & \multicolumn{1}{c|}{\textcolor{green}{\huge \ding{51}}} & \multicolumn{1}{c|}{} & \multicolumn{1}{c|}{} & \multicolumn{1}{c|}{} & \multicolumn{1}{c|}{} &  & \multicolumn{1}{c|}{Function} & \textcolor{red}{\huge \ding{55}} & \multicolumn{1}{c|}{\textcolor{red}{\huge \ding{55}}} & \textcolor{red}{\huge \ding{55}} \\ \hline
Wu et al.~\cite{DBLP:conf/issta/WuJPLD0BS23} & \begin{tabular}[c]{@{}c@{}}\textless{}1B,\\ 1-10B,\\ \textgreater{}10B\end{tabular} & \multicolumn{1}{c|}{} & \multicolumn{1}{c|}{} & \multicolumn{1}{c|}{} & \multicolumn{1}{c|}{} & \multicolumn{1}{c|}{} & \textcolor{green}{\huge \ding{51}} & \multicolumn{1}{c|}{Function} & \textcolor{green}{\huge \ding{51}} & \multicolumn{1}{c|}{\textcolor{red}{\huge \ding{55}}} & \textcolor{red}{\huge \ding{55}} \\ \hline
Fu et al.~\cite{fu2023chatgpt} & \textgreater{}10B & \multicolumn{1}{c|}{} & \multicolumn{1}{c|}{} & \multicolumn{1}{c|}{} & \multicolumn{1}{c|}{} & \multicolumn{1}{c|}{} & \textcolor{green}{\huge \ding{51}} & \multicolumn{1}{c|}{Function} & \textcolor{red}{\huge \ding{55}} & \multicolumn{1}{c|}{\textcolor{red}{\huge \ding{55}}} & \textcolor{red}{\huge \ding{55}} \\ \hline
Zhang et al.~\cite{zhang2023pre}  & \textless{}1B & \multicolumn{1}{c|}{} & \multicolumn{1}{c|}{} & \multicolumn{1}{c|}{\textcolor{green}{\huge \ding{51}}} & \multicolumn{1}{c|}{} & \multicolumn{1}{c|}{} &  & \multicolumn{1}{c|}{Function} & \textcolor{red}{\huge \ding{55}} & \multicolumn{1}{c|}{\textcolor{red}{\huge \ding{55}}} & \textcolor{red}{\huge \ding{55}} \\ \hline
Pearce et al.~\cite{pearce2023examining} & \begin{tabular}[c]{@{}c@{}}\textless{}1B,\\ 1-10B,\\ \textgreater{}10B\end{tabular} & \multicolumn{1}{c|}{} & \multicolumn{1}{c|}{} & \multicolumn{1}{c|}{} & \multicolumn{1}{c|}{} & \multicolumn{1}{c|}{} & \textcolor{green}{\huge \ding{51}} & \multicolumn{1}{c|}{Function} & \textcolor{green}{\huge \ding{51}} & \multicolumn{1}{c|}{\textcolor{red}{\huge \ding{55}}} & \textcolor{red}{\huge \ding{55}} \\ \hline
Wei et al.~\cite{DBLP:journals/ejwcn/WeiBWLYSL23} & \textless{}1B & \multicolumn{1}{c|}{\textcolor{green}{\huge \ding{51}}} & \multicolumn{1}{c|}{} & \multicolumn{1}{c|}{} & \multicolumn{1}{c|}{} & \multicolumn{1}{c|}{} &  & \multicolumn{1}{c|}{Function} & \textcolor{red}{\huge \ding{55}} & \multicolumn{1}{c|}{\textcolor{red}{\huge \ding{55}}} & \textcolor{red}{\huge \ding{55}} \\ \hline
Fu et al.~\cite{fu2023vision} & \textless{}1B & \multicolumn{1}{c|}{} & \multicolumn{1}{c|}{\textcolor{green}{\huge \ding{51}}} & \multicolumn{1}{c|}{} & \multicolumn{1}{c|}{} & \multicolumn{1}{c|}{} &  & \multicolumn{1}{c|}{Function} & \textcolor{red}{\huge \ding{55}} & \multicolumn{1}{c|}{\textcolor{red}{\huge \ding{55}}} & \textcolor{red}{\huge \ding{55}} \\ \hline
Islam et al.~\cite{islam2024code} & \multicolumn{1}{l|}{1-10B} & \multicolumn{1}{c|}{} & \multicolumn{1}{c|}{} & \multicolumn{1}{c|}{} & \multicolumn{1}{c|}{\textcolor{green}{\huge \ding{51}}} & \multicolumn{1}{c|}{} &  & \multicolumn{1}{c|}{Function} & \textcolor{red}{\huge \ding{55}} & \multicolumn{1}{c|}{\textcolor{red}{\huge \ding{55}}} & \textcolor{red}{\huge \ding{55}} \\ \hline
Fu et al.~\cite{DBLP:journals/ese/FuTLKNPG24} & \textless{}1B & \multicolumn{1}{c|}{} & \multicolumn{1}{c|}{} & \multicolumn{1}{c|}{} & \multicolumn{1}{c|}{} & \multicolumn{1}{c|}{\textcolor{green}{\huge \ding{51}}} &  & \multicolumn{1}{c|}{Function} & \textcolor{red}{\huge \ding{55}} & \multicolumn{1}{c|}{\textcolor{red}{\huge \ding{55}}} & \textcolor{red}{\huge \ding{55}} \\ \hline
Zhou et al.~\cite {zhou2024out} & \textless{}1B & \multicolumn{1}{c|}{\textcolor{green}{\huge \ding{51}}} & \multicolumn{1}{c|}{} & \multicolumn{1}{c|}{\textcolor{green}{\huge \ding{51}}} & \multicolumn{1}{c|}{} & \multicolumn{1}{c|}{} &  & \multicolumn{1}{c|}{Function} & \textcolor{red}{\huge \ding{55}} & \multicolumn{1}{c|}{\textcolor{red}{\huge \ding{55}}} & \textcolor{red}{\huge \ding{55}} \\ \hline
Nong et al.~\cite{DBLP:journals/corr/abs-2402-17230} & \textgreater{}1B & \multicolumn{1}{c|}{} & \multicolumn{1}{c|}{} & \multicolumn{1}{c|}{} & \multicolumn{1}{c|}{} & \multicolumn{1}{c|}{} & \textcolor{green}{\huge \ding{51}} & \multicolumn{1}{c|}{Function} & \textcolor{red}{\huge \ding{55}} & \multicolumn{1}{c|}{\textcolor{red}{\huge \ding{55}}} & \textcolor{red}{\huge \ding{55}} \\ \hline
Berabi et al.~\cite{DBLP:journals/corr/abs-2402-13291} & \textgreater{}1B & \multicolumn{1}{c|}{\textcolor{green}{\huge \ding{51}}} & \multicolumn{1}{c|}{} & \multicolumn{1}{c|}{} & \multicolumn{1}{c|}{} & \multicolumn{1}{c|}{} &  & \multicolumn{1}{c|}{Function} & \textcolor{red}{\huge \ding{55}} & \multicolumn{1}{c|}{\textcolor{red}{\huge \ding{55}}} & \textcolor{red}{\huge \ding{55}} \\ \hline
Ahmad et al.~\cite{DBLP:journals/corr/abs-2302-01215} & \textgreater{}10B & \multicolumn{1}{c|}{} & \multicolumn{1}{c|}{} & \multicolumn{1}{c|}{} & \multicolumn{1}{c|}{} & \multicolumn{1}{c|}{} & \textcolor{green}{\huge \ding{51}} & \multicolumn{1}{c|}{Function} & \textcolor{red}{\huge \ding{55}} & \multicolumn{1}{c|}{\textcolor{red}{\huge \ding{55}}} & \textcolor{red}{\huge \ding{55}} \\ \hline
Islam et al.~\cite{DBLP:journals/corr/abs-2401-03374} & \textless{}10B & \multicolumn{1}{c|}{} & \multicolumn{1}{c|}{\textcolor{green}{\huge \ding{51}}} & \multicolumn{1}{c|}{} & \multicolumn{1}{c|}{} & \multicolumn{1}{c|}{} &  & \multicolumn{1}{c|}{Function} & \textcolor{red}{\huge \ding{55}} & \multicolumn{1}{c|}{\textcolor{green}{\huge \ding{51}}} & \textcolor{red}{\huge \ding{55}} \\ \hline
Nong et al.~\cite{nong2024automated} & >10B & \multicolumn{1}{c|}{} & \multicolumn{1}{c|}{} & \multicolumn{1}{c|}{} & \multicolumn{1}{c|}{} & \multicolumn{1}{c|}{} & \multicolumn{1}{c|}{\textcolor{green}{\huge \ding{51}}}  & \multicolumn{1}{c|}{Function} & \textcolor{red}{\huge\ding{55}} & \multicolumn{1}{c|}{\textcolor{red}{\huge \ding{55}}} & \textcolor{red}{\huge \ding{55}} \\ \hline
Wang et al.~\cite{DBLP:journals/corr/abs-2405-04994} & >1B & \multicolumn{1}{c|}{} & \multicolumn{1}{c|}{} & \multicolumn{1}{c|}{} & \multicolumn{1}{c|}{} & \multicolumn{1}{c|}{} &  \multicolumn{1}{c|}{\textcolor{green}{\huge \ding{51}}} & \multicolumn{1}{c|}{Function} & \textcolor{red}{\huge \ding{55}} & \multicolumn{1}{c|}{\textcolor{red}{\huge \ding{55}}} & \textcolor{red}{\huge \ding{55}} \\ 
\bottomrule
\end{tabular}
}
\label{table4repair}
\end{table*}

\vspace{0.2cm}
\noindent\textbf{Domain-specific Pre-training.}
Domain-specific pre-training involves pre-training an LLM on data specific to a particular domain before fine-tuning it for the target task. 
Considering the similarity between bug-fixing and vulnerability-fixing tasks, 
several studies~\cite{DBLP:journals/tse/ChenKM23,DBLP:journals/tse/ChiQLZY23,zhang2023pre,zhou2024out} considered the task of bug-fixing as a pre-training task to enhance LLMs. Specifically, they first pre-trained LLMs on a bug fix corpus to fix bugs, and then fine-tuned LLMs on a vulnerability fix dataset to repair vulnerabilities. This kind of pre-training technique can be considered as \textit{Transfer Learning}.

\vspace{0.2cm}
\noindent\textbf{Reinforcement Learning.}
Islam et al.~\cite{islam2024code} introduced SecureCode, an LLM tuned using a reinforcement learning framework. This approach integrated syntactic and semantic rewards to generate fixes for vulnerable code. 
Specifically, they leveraged the CodeBLEU~\cite{DBLP:journals/corr/abs-2009-10297} score as the syntactic reward and BERTScore~\cite{DBLP:conf/iclr/ZhangKWWA20} as the semantic reward. 
After combining these rewards, they applied the Proximal Policy Optimization (PPO) algorithm~\cite{schulman2017proximal} to fine-tune the CodeGen2-7B~\cite{nijkamp2023codegen2} model.

\subsection{\textbf{Prompt Engineering.}}
\label{sec_repair:PromptEngineering}

\vspace{0.2cm}
\noindent\textbf{Zero-shot Prompting.}
For zero-shot prompting, researchers have sought to create effective prompts that guide LLMs in performing vulnerability repair. Prompt engineering typically includes components outlined below:

\begin{itemize}
    \item \textit{Vulnerability Descriptions.} This component provides specific details about the types of vulnerabilities to be addressed, helping the LLM understand the nature of the issues within the code. Specifically, Pearce et al.~\cite{pearce2023examining} propose to use a format like ``// BUG: [vulnerability-description]'' to provide the descriptions on the vulnerability. 
    Ahmad et al.~\cite{DBLP:journals/corr/abs-2302-01215} provided the bug description with explicit instructions for repair with LLMs, as illustrated in the prompt: ``// BUG: Access Control Check Implemented After the Asset is Accessed. // Ensure that access is granted before data is accessed''.
    
    \item \textit{Vulnerability Location.} This element indicates the specific sections of the code where vulnerabilities may exist. Wu et al.~\cite{DBLP:conf/issta/WuJPLD0BS23} proposed a prompt that begins by commenting on the buggy code lines and adds "BUG:" to highlight them, which provides the location of the vulnerability to LLMs. 
    
    \item \textit{Vulnerability-Related Auxiliary Information.} This component encompasses auxiliary information about vulnerabilities. Nong et al.~\cite{DBLP:journals/corr/abs-2402-17230} incorporate vulnerability semantics—specifically, the behaviors of a vulnerable program that contribute to its susceptibility—into the prompt to enhance the performance of LLMs.

    \item \textit{Program-Analysis Auxiliary Information.}
    Wang et al.~\cite{DBLP:journals/corr/abs-2405-04994} proposed a program-analysis-based approach. Given a patch, their method first utilizes both the vulnerable code and its fixed version as input. The process begins by localizing the Minimum Edit Node (MEN), which is the common ancestor node of all modified nodes. Once the MEN is identified, customized rules are applied based on the MEN type to extract vulnerable code patterns. These patterns are then incorporated into the prompt to assist LLMs in generating more effective vulnerability fixes.

\end{itemize}

\vspace{0.2cm}
\noindent\textbf{Few-shot Prompting.}
Furthermore, for few-shot prompting, 
Fu et al.~\cite{fu2023chatgpt} investigated a straightforward prompt strategy involving the provision of three repair examples within each prompt. This approach aimed to aid GPT-3.5 and GPT-4 in performing the repair task more effectively.
Nong et al.~\cite{nong2024automated} proposed a novel few-shot prompting method. Given a vulnerable program with a known vulnerability location, they first narrowed the analysis scope to only the relevant subset of the program. Next, they elicited the LLM to identify the vulnerability’s root cause within this reduced scope, dynamically selecting exemplars from a pre-mined database that best fit the program.  With the chosen exemplars, they formed the patching prompt to generate patches with LLMs.

\begin{tcolorbox}[left=4pt,right=4pt,top=2pt,bottom=2pt,boxrule=0.5pt]
\textbf{Answer to RQ3:} 
Our analysis revealed three commonly used techniques to adapt LLMs for vulnerability repair: including \emph{fine-tuning} ($\approx$63\%) and \emph{ prompt engineering} ($\approx$37\%).
\end{tcolorbox}

\section{RQ4: What are the characteristics of datasets and deployment?}\label{RQ4}

A typical learning-based pipeline can be divided into three phases: 1) data preparation, 2) model construction, and 3) deployment. The previous RQs primarily focus on the model construction phase. In this RQ, we focus on examining the characteristics of the datasets used, as well as the deployment strategies employed in LLM-based vulnerability detection and repair studies.

\subsection{\mrevised{Data Characteristics}}
\label{DataCharacteristics}

In our investigation of the data utilized in LLM-based vulnerability detection and repair studies, we primarily focus on two aspects: 1) input code size and 2) data quality.

\subsubsection{\mrevised{Input Granularity}}
Input granularity refers to the size or scope of the code provided to the model for vulnerability detection and repair. For example, in a vulnerability detection task with repo-level input granularity, the entire repository is given to the model, and it is tasked with identifying vulnerabilities within that repository. Input granularity can vary from the line-level, function-level, and class-level to repo-level.

For vulnerability detection, as shown in Table~\ref{table4detection}, the majority of studies concentrate on function-level vulnerability detection, with only 7 addressing line-level detection. The function-level approach analyzes entire functions for potential vulnerabilities, whereas line-level detection examines specific lines of code, which can be advantageous for pinpointing fine-grained locations of vulnerabilities. Regarding vulnerability repair, Table~\ref{table4repair} reveals that all studies target function-level repairs.

However, there is a notable gap in research exploring larger input granularity for both vulnerability detection and repairs, such as class-level and repo-level approaches. Larger input granularity is essential for identifying vulnerabilities that span multiple functions. Specifically, there are no studies focused on class-level vulnerability detection and repair. For repo-level detection, one recent work~\cite{zhou2024comparison} initially explored the use of LLMs (e.g., Code Llama) to detect vulnerabilities across entire repositories, comparing their performance with SAST tools (e.g., CodeQL).
Notably, none of the studied papers investigate repo-level vulnerability repair. Overall, research on vulnerability detection and repair for larger input granularity remains limited.

\subsubsection{Data Quality}
We separately investigate the data quality issues in vulnerability detection and repair. Specifically, we find that vulnerability detection studies primarily utilize datasets with labels generated by heuristic-based methods. Additionally, vulnerability repair studies often rely on datasets that lack test cases, making it difficult to fully evaluate the correctness of the generated repairs since there can be multiple correct fixes.

\vspace{0.2cm}
\noindent\textbf{Vulnerability detection studies mainly rely on datasets using heuristic-based labeling methods.}
Table~\ref{table4detection} shows that many studies have employed heuristic-based labeling functions to generate data for vulnerability detection tasks. Typically, they collect vulnerability-fixing commits from databases like NVD and label the pre-commit versions of the functions modified by these commits as vulnerable. Additionally, all unchanged functions in the same files are labeled as non-vulnerable. In this way, the vulnerable functions are those in the files affected by vulnerability-fixing commits.
However, recent studies~\cite{DBLP:conf/icse/CroftBK23,DBLP:conf/issta/NieLWWLW23,DBLP:conf/raid/0001DACW23} have highlighted quality issues with vulnerability data generated by the heuristic-based method, including noisy or incorrect labels (e.g.,  labeling clean code as vulnerable).

\vspace{0.2cm}
\noindent\textbf{Vulnerability repair studies mainly rely on datasets without test cases.}
As present in Table~\ref{table4repair}, obtaining real-world vulnerability data along with corresponding test cases is challenging. Most studies either utilize synthetic vulnerable code and the corresponding fixes (e.g., ~\cite{islam2024code}) or use real-world vulnerable code and fixes without test cases (e.g., ~\cite{DBLP:journals/tse/ChenKM23,DBLP:journals/tse/ChiQLZY23,DBLP:conf/sigsoft/FuTLNP22}). 
Two recent studies~\cite{DBLP:conf/issta/WuJPLD0BS23, pearce2023examining} utilized real-world vulnerability-fixing datasets with test cases, however, the number of samples is very limited, with only 42 and 12 samples being evaluated, respectively.
This underscores the urgent need for more extensive real-world datasets and accompanying test cases.

\subsection{Deployment Strategies}\label{DeploymentStrategies}

In our investigation of the deployment strategies utilized in LLM-based vulnerability detection and repair studies, we focus on two key aspects: 1) the communication and collaboration between developers and LLM-based methods, and 2) the integration of these methods into developers' workflows.
We focus on the communication and collaboration between users and LLM-based methods because they are essential in deploying LLMs in practice and are pivotal in maximizing the real-world value of LLMs~\cite{DBLP:journals/tkdd/YangJTHFJZYH24}. Moreover, a recent survey~\cite{DBLP:conf/icse/Liang0M24} shows that developers desire opportunities for natural language interaction with LLM-based solutions, also highlighting the significance of this capability.
In addition, seamless integration of these solutions into developers' workflows is crucial for successful deployment. A recent study~\cite{DBLP:conf/seke/Zhang0ZAW23} found that even for the popular LLM-based tool GitHub Copilot, the top 1 limitation reported is the difficulty of integrating Copilot with IDEs or other plugins. This emphasizes the need for better support in integrating LLM-based tools with other frequently used tools in developers' workflows such as IDEs.

Firstly, as shown in Table~\ref{table4detection}, limited studies on vulnerability detection in our review support communication or collaboration between developers and LLM-based methods. One exception is a study~\cite{yang2024security} where the method generates explanations for vulnerabilities during the detection process.
However, interaction and communication with developers are essential; for instance, unexplained suggestions may be disregarded, as developers need to understand the rationale behind the recommendations to trust and implement them effectively. In vulnerability repair studies (as shown in Table~\ref{table4repair}), most also do not support communication between developers. An exception is a recent study by Islam et al.~\cite{DBLP:journals/corr/abs-2401-03374}, which provides explanations for the fixes when addressing vulnerabilities.

Secondly, as illustrated in Table~\ref{table4detection} and Table~\ref{table4repair}, all studies in our review are not integrated into developers' workflows and tools but are evaluated only offline using static and historical data. For these solutions to be effectively adopted, they should seamlessly integrate into existing development environments and tools.

\begin{tcolorbox}[left=4pt,right=4pt,top=2pt,bottom=2pt,boxrule=0.5pt]
\textbf{Answer to RQ4:} 
The datasets used in LLM-based vulnerability detection and repair studies primarily focus on function-level or line-level inputs, with limited exploration of class or repository level. Many vulnerability detection datasets employ heuristic-based labeling methods, and repair datasets frequently lack associated test cases. Furthermore, current approaches do not emphasize integration with developer workflows or interaction and collaboration between developers and LLM-based models.
\end{tcolorbox}

\section{The Road Ahead}\label{ChaAndOpp}
In this section, we first discuss the limitations of current studies and then introduce a roadmap, illustrating the research trajectory shaped by prior work and highlighting future directions for exploration.

\subsection{Limitations}\label{Challenges}

\vspace{0.2cm}
\noindent\textbf{Limitation 1: Small Input Granularity.}
Currently, LLM-based vulnerability detection and repair solutions primarily target the function level (or more fine-grained, at the line level). However, this small input granularity indicates that these approaches may not perform optimally when presented with a wider range of programs, such as classes or a whole repository.
Function-level vulnerability detection approaches could overlook vulnerabilities that span multiple functions or classes~\cite{sejfia2024toward}. Similarly, function-level vulnerability repair approaches fall short when tasked with modifications across multiple functions within the repository~\cite{zhou2024out}.
\textit{In addition to functions, future research could propose LLM-based detection/repair approaches capable of handling a broader range of programs.}

\vspace{0.2cm}
\noindent\textbf{Limitation 2: Lack of High-quality Vulnerability Dataset.}
One major challenge is the lack of high-quality vulnerability datasets. For vulnerability repair, previous studies~\cite{DBLP:conf/icse/CroftBK23,DBLP:conf/issta/NieLWWLW23,DBLP:conf/raid/0001DACW23} have highlighted issues with existing vulnerability data, including noisy or incorrect labels (e.g.,  labeling clean code as vulnerable). 
This data quality issue is primarily attributed to the use of automatic vulnerability collection~\cite{DBLP:conf/issta/NieLWWLW23}, which can gather large enough data for training DL-based models including LLMs but cannot ensure the complete correctness of the labels. While manually checking each data sample can ensure high quality, it is a very tedious and expensive process, especially when aiming for a large dataset. \textit{Constructing a high-quality vulnerability detection benchmark remains an open challenge to date.}

For vulnerability repair, most studies either utilize synthetic data (e.g., ~\cite{islam2024code}) or use real-world data without test cases (e.g., ~\cite{DBLP:journals/tse/ChenKM23,DBLP:journals/tse/ChiQLZY23,DBLP:conf/sigsoft/FuTLNP22}). The reliance on synthetic data can limit the generalizability of the findings, as it may not accurately reflect the complexities and nuances of actual vulnerabilities encountered in production environments. 
On the other hand, real-world data without test cases makes it difficult to evaluate the effectiveness of proposed solutions comprehensively, as the absence of test cases hampers the ability to verify the correctness of the vulnerability repairs. 
\textit{There is an urgent need for large real-world vulnerability-fixing datasets that include test cases.}

Moreover, there is a growing concern that the capabilities of LLMs may derive from the inclusion of evaluation datasets in the pre-training corpus of LLMs, a phenomenon known as data contamination~\cite{jiang2024investigating}. To mitigate this concern, high-quality vulnerability detection/fixing benchmarks are preferred to have no overlap with the pre-training corpus of LLMs.

\vspace{0.2cm}
\noindent\textbf{Limitation 3: Suboptimal Performance Caused by Complexity in Vulnerability Data.}
Vulnerabilities can be inherently complicated, which brings challenges for their detection and repair with LLMs.
For instance, inter-procedural vulnerabilities are prevalent in vulnerability data, and Li et al.~\cite{li2024effectiveness} discovered that detecting inter-procedural vulnerabilities poses greater challenges than intra-procedural ones.
Moreover, vulnerabilities encompass a wide array of Common Weakness Enumeration (CWE) types~\cite{zhou2023devil}, but LLMs may struggle with less frequent CWE types compared to frequent types~\cite{zhou2024out,zhou2023devil}.
In addition, vulnerabilities are typically represented in terms of code units, such as code lines, functions, or program slices within which the vulnerabilities occur. Sejfia et al.~\cite{sejfia2024toward} observed a significant accuracy drop when detecting vulnerabilities that span multiple code units, such as spanning multiple functions. \textit{Future research should consider the complex nature of vulnerabilities when designing LLM-based solutions.}

\vspace{0.2cm}
\noindent\textbf{Limitation 4: Reliance on Lightweight LLMs.}
As highlighted in Table~\ref{table4detection} and Table~\ref{table4repair}, most studies adapted lightweight LLMs (\#parameter < 1B) for vulnerability detection and repair. 
For lightweight LLMs like CodeBERT, researchers have explored various strategies to boost their performance, including data-centric enhancements, model-centric innovations, integration with program analysis, combining LLMs with other deep learning methods, domain-specific pre-training, causal learning, and reinforcement learning.
In some cases, such as Liu et al.~\cite{liu2024pre}, multiple techniques have been combined to achieve even greater performance, highlighting the versatility and potential of lightweight models.
In contrast, the use of large LLMs (with over 1 billion parameters) remains relatively limited compared to lightweight LLMs, leaving a large room to explore.

\vspace{0.2cm}
\noindent\textbf{Limitation 5: Lack of Deployment Consideration.}
We examine two crucial aspects of deployment: 1) interaction with developers and 2) integration into developers' current workflows.
Currently, none of the studies in our review have been incorporated into developers' workflows; instead, they primarily rely on static and historical data to evaluate their effectiveness.
Moreover, limited studies in our review incorporate interaction with developers, such as engaging with developers through feedback or explaining the rationale behind vulnerability detection. 
Limited interaction between developers and LLM-based solutions may impede the establishment of trust and collaboration during practical applications. This lack of interactive features restricts the practical applicability of these models in real-world scenarios, where effective vulnerability management relies on collaboration between automated tools and developers.

To address these gaps, future research should explore more effective strategies for fostering collaboration and trust between developers and LLM-based solutions~\cite{DBLP:journals/corr/abs-2309-04142}. Additionally, seamlessly integrating LLM-based solutions can enable them to evolve into intelligent partners, providing enhanced support to developers.

\vspace{0.2cm}
\noindent\textbf{Limitation 6: Lack of High Accuracy and Robustness.}
A vulnerability detection or repair solution with high accuracy is generally preferred, as it boosts developers' confidence in the reliability of detections and fixes. 
However, current state-of-the-art approaches~\cite{liu2024pre, zhou2024out} have not yet achieved satisfactory accuracy, with 67.6\% and 20\% accuracy scores for vulnerability detection and repair, respectively.
Moreover, the solution should maintain robustness against data perturbations or adversarial attacks to ensure its resilience.
However, Yang et al.~\cite{yang2022natural} and Rahman et al.~\cite{mahbubur2023towards} discovered that LLMs are not robust against data perturbations. \textit{Future research should seek ways to improve the accuracy and robustness of LLM-based solutions.}

\begin{figure*}[t]
  \centering
  \includegraphics[width=0.95\linewidth]{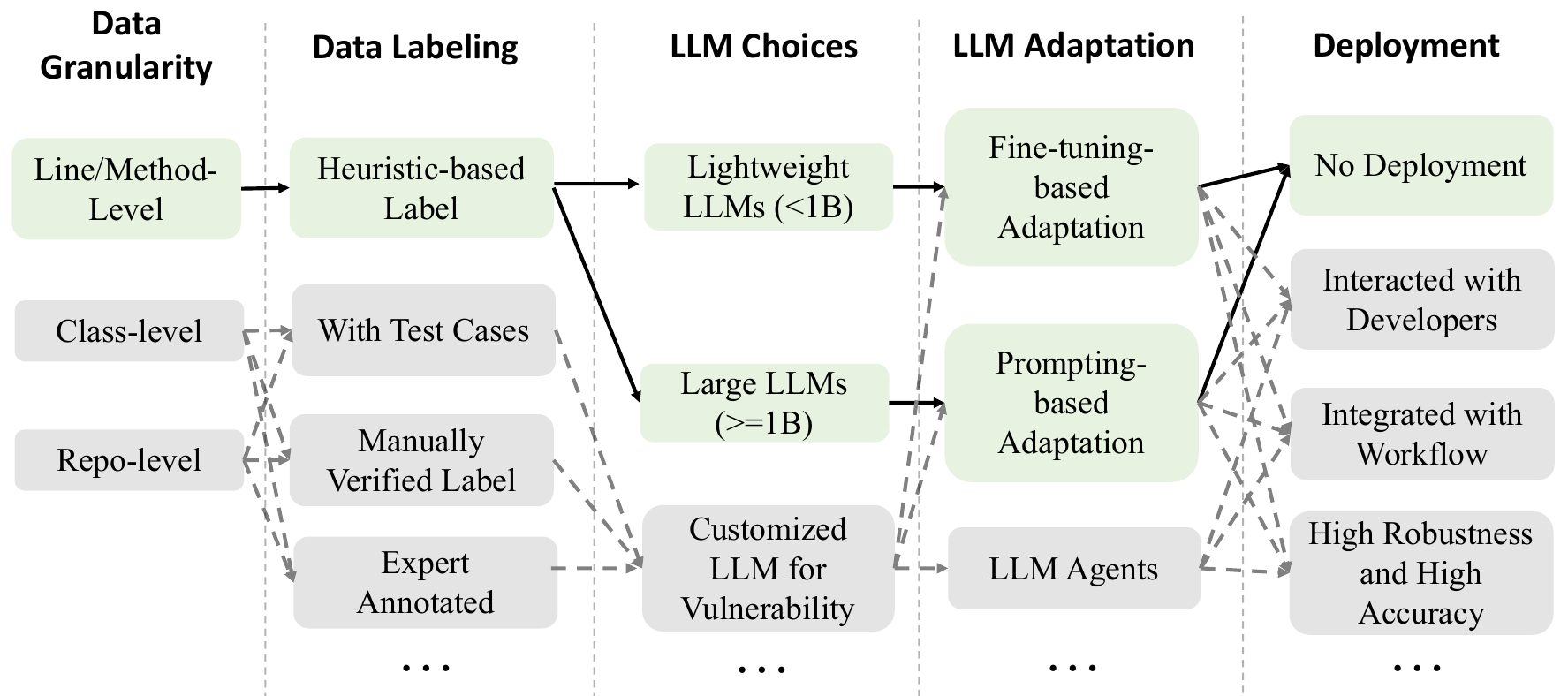}
\caption{Roadmap and Future Directions}
\label{fig:roadmap}
\end{figure*}

\subsection{Roadmap}\label{Roadmap}

Figure~\ref{fig:roadmap} presents the roadmap. The green modules highlight the aspects that existing studies have primarily focused on, while the gray modules represent areas that have not been extensively explored but warrant further investigation. The solid black lines represent the trajectories followed by most existing LLM-based methods for vulnerability detection and repair, while the gray dotted lines highlight relatively under-explored paths that could offer promising directions for future research in vulnerability detection and repair.

\subsubsection{Established Pathways in Current Studies.}
As shown in both Table~\ref{table4detection}--\ref{table4repair} and the solid black lines in Figure~\ref{fig:roadmap}, the existing methods typically leverage either lightweight LLMs or large LLMs adapted using techniques such as fine-tuning or prompting. The existing methods often focus on method-level or line-level analysis. The datasets used are usually collected from heuristic approaches for vulnerability detection or real-world data without test cases for vulnerability repair. Finally, these methods usually neglect deployment considerations, such as ensuring effective interaction with developers.

\subsubsection{Path Forward.}
As shown in Figure~\ref{fig:roadmap}, there are many promising yet under-explored modules (highlighted in green) and paths (represented by the gray dotted lines.) Looking ahead, we can outline three progressively advanced stages for exploring new directions.
In the first stage, researchers can study these under-explored modules individually. In the second stage, they can investigate the under-explored paths, which may include combinations of under-explored modules and well-established ones or consist solely of multiple under-explored modules.
Finally, in the third stage, by transforming all under-explored modules into well-understood components and studying various under-explored paths and combinations, researchers can summarize their findings and analyze strategies for better developing LLM-based solutions, ultimately leading to the new generation of powerful LLM-based vulnerability detection and repair solutions. Below, we describe the three stages in detail.

\vspace{0.2cm}
\noindent\textbf{Stage 1: Exploring Under-explored Modules Individually.}
Figure~\ref{fig:roadmap} highlights several under-explored modules in green, each offering exciting potential for discovery. We outline below some promising opportunities for deeper investigation into these areas:

\begin{itemize}[leftmargin=*]

 \item \textit{Opportunity 1: Curating High-quality Test Set for Vulnerability Detection.}
The absence of a high-quality vulnerability dataset poses a significant obstacle to vulnerability detection. While obtaining fully correct labels for a large dataset is expensive, a viable solution is to curate a high-quality test set (which is much smaller than the whole dataset) that can accurately assess progress in vulnerability detection.
An easy approach for future work is to combine the manually checked vulnerability data samples scattered across several separate research works (e.g.,~\cite{DBLP:conf/icse/CroftBK23,he2023large,li2023comparison}) to form a high-quality test set. Future works can then do an empirical study to recognize the real progress in vulnerability detection with the high-quality test set. Furthermore, the community can maintain a living high-quality test set by adding new manually verified data to it. This curated test set can serve as a reliable benchmark for vulnerability detection. 

\vspace{0.2cm}
\item \textit{Opportunity 2: Repo-level Vulnerability Detection/Repair.}
Current vulnerability detection and repair techniques primarily focus on the function or line level. One key reason is the input length limitations (512 subtokens) of the small LLMs like CodeBERT and CodeT5, which are predominantly used in existing studies. The 512 subtokens limitation aligns well with the function level data but faces difficulty in scaling up to classes or repositories.
However, the emergence of recent larger LLMs with significantly higher input length capacities, such as GPT-4 (which can handle 128k subtokens), facilitates more effective processing of repo-level data. This presents an opportunity for future research to explore repo-level vulnerability detection/repair by leveraging these larger LLMs.
For this direction, one recent study~\cite{zhou2024comparison} has initially explored the repo-level vulnerability detection task. Specifically, it compared SAST tools (e.g., CodeQL) with popular or state-of-the-art open-source LLMs (e.g., Code Llama) for detecting software vulnerabilities in software repositories. The experimental results indicated that SAST tools achieve low vulnerability detection rates with relatively low false positives, whereas LLMs can detect more vulnerabilities but tend to suffer from high false positive rates. Alongside this initial effort, there is considerable potential and opportunities for further research in this direction.

\vspace{0.2cm}
\item \textit{Opportunity 3: {Customized LLMs for Vulnerability.}}
Currently, widely used LLMs in vulnerability detection/repair are general-purpose LLMs (e.g., CodeBERT, CodeT5, and GPT-3.5) that do not fully exploit the wealth of open-source vulnerability data. 
A promising avenue is the development of customized LLMs tailored for vulnerability data. Some initial attempts in this direction include vulnGPT~\cite{vul_GPT} and Microsoft Security Copilot~\cite{SecurityCopilot}. However, as these solutions are proprietary, their customized LLM details may not be fully disclosed. We advocate for collaborative efforts to develop open-sourced and effective customized LLMs for vulnerability.

\vspace{0.2cm}
\item \textit{Opportunity 4: Advanced LLM Usage and Adaptation.}
Regarding LLMs usages, beyond the techniques observed in the included studies—such as \emph{fine-tuning}, \emph{prompt engineering}, and \emph{retrieval augmentation}—there exist two more advanced usages of LLMs~\cite{minaee2024large} that have not been explored yet in vulnerability detection and repair: 1) \emph{LLM Agent}: LLMs can serve as agents to decompose complex tasks into smaller components and employ multiple LLMs to address them~\cite{wang2024survey}; 2) \emph{Usage of External Tools}: LLMs can utilize external tools such as search engines, external databases, and other resources to enhance them~\cite{qin2023toolllm}. 
Regarding detailed LLMs adaptation techniques, as illustrated in Fig.\ref{ExpTec} and Fig.\ref{ExpTec2}, the majority of proposed adaptations in this field are designed for \emph{fine-tuning} and \emph{prompt engineering}. However, a plethora of advanced adaptation techniques for \emph{retrieval augmentation} remains under-explored. 
These techniques include iterative retrieval augmentation~\cite{shao2023enhancing} and recursive retrieval augmentation~\cite{trivedi2022interleaving}, and adaptive retrieval augmentation~\cite{asai2023self}. 
Researchers can consider using those unexplored advanced LLM usages/adaptations in future works.

\vspace{0.2cm}
\item \textit{Opportunity 5: Support Deployment-ready Features.} Efforts could focus on enhancing LLM-based solutions with deployment-ready features, such as user interaction capabilities and seamless integration into existing developer workflows. This includes developing intuitive interfaces that allow developers to provide real-time feedback on suggestions made by the LLMs, thereby creating a more collaborative environment.
Additionally, implementing functionalities that explain the reasoning behind the model's recommendations can help demystify the decision-making process, fostering trust and facilitating better adoption among developers.
Moreover, integrating these solutions into popular IDEs or version control systems can streamline the workflow, making it easier for developers to utilize LLMs as part of their daily practices. By prioritizing these enhancements, LLM-based tools can significantly improve their usability and effectiveness in real-world scenarios, ultimately contributing to more robust vulnerability detection and repair processes.

\end{itemize}

\vspace{0.2cm}
\noindent\textbf{Stage 2: Exploring Under-explored Paths.}
The under-explored paths (mainly represented by gray dotted lines from data to deployment) may include combinations of under-explored modules with well-established ones or consist solely of multiple under-explored modules. This stage presents numerous opportunities. Below, we first highlight some examples of studying these combinations of under-explored modules alongside well-established ones:

\begin{itemize}[leftmargin=*]
    \item Once the community develops customized LLMs, researchers can apply well-studied fine-tuning or prompting-based adaptation techniques to improve function-level or line-level vulnerability detection or repair. These enhanced models can further be extended to support developer interaction, creating more practical and integrated solutions.

    \item Once new datasets are established—such as those containing manually verified vulnerabilities and associated test cases—researchers can re-evaluate existing LLM methods to assess their effectiveness in vulnerability detection and repair. Exploring various combinations of under-explored aspects and well-studied approaches can yield valuable insights and advancements.

    \item Researchers can leverage general-purpose LLMs (e.g., CodeBERT, CodeT5, and GPT-3.5) to develop repo-level vulnerability detection and repair tools, also facilitating interaction with developers.

\end{itemize}

There are also many opportunities in paths that consist solely of multiple under-explored modules. For instance, researchers could design a repo-level vulnerability detection or repair method using customized vulnerability-specific LLMs and enhance its effectiveness by incorporating LLM agent techniques. Finally, the tool can be integrated into IDEs, leveraging LLMs to improve interaction capabilities.

\vspace{0.2cm}
\noindent\textbf{Stage 3: Summarizing Findings and Applying Them for Next-Generation LLM-based Solutions.}
In this most advanced stage, researchers are expected to transform various under-explored aspects into well-studied areas and investigate diverse under-explored paths. At this stage, they can summarize their findings and analyze strategies for improving the development of LLM-based solutions. By leveraging this experience and knowledge, they may push the boundaries of what LLMs can achieve in vulnerability detection and repair to new heights.

Ideally, by 2030, following the roadmap illustrated in Figure~\ref{fig:roadmap}, the community can develop a highly effective method capable of performing line-level, method-level, class-level, and repository-level vulnerability detection and repair, utilizing an expert-annotated benchmark to ensure accurate evaluation. Furthermore, the ideal solution should facilitate seamless communication and collaboration with developers, achieving high accuracy and robustness while providing trustworthy, expert-level insights into real-world vulnerabilities.

\section{Threats to Validity}\label{limitation}
The potential threat to validity is the risk of inadvertently excluding relevant studies during the literature search and selection phase. Incomplete summarization of keywords for vulnerability detection/repair or varied terminologies of LLMs may have caused relevant research studies to be missed in our review. To mitigate this risk, we initially performed a manual selection of 17 high-impact venues and extracted a relatively comprehensive set of standard keywords from relevant papers within these venues. In addition, we further augmented our search results by combining automated search with forward-backward snowballing.

\section{Conclusion and Future Work}\label{Conclusion}
The use of Large Language Models (LLMs) for vulnerability detection and repair has been garnering increasing attention. This paper presents a systematic literature review of 58 primary studies on LLMs for vulnerability detection and repair.
This review begins by analyzing the types of LLMs used in primary studies, shedding light on researchers' preferences for different LLMs. Subsequently, we categorized a variety of techniques for adapting LLMs.
Through our analysis, this review also identifies the limitations in this field and proposes a research roadmap outlining promising avenues for future exploration. 
In the future, we plan to broaden this literature review by incorporating additional vulnerability-related tasks, such as vulnerability localization and vulnerability assessment.

\vspace{0.2cm}
\noindent \textbf{Acknowledgement.}  This research / project is supported by the National Research Foundation, under its Investigatorship Grant (NRF-NRFI08-2022-0002). Any opinions, findings and conclusions or recommendations expressed in this material are those of the author(s) and do not reflect the views of National Research Foundation, Singapore.

\balance
\bibliographystyle{acm}
\bibliography{main}

\end{document}